\documentclass[12pt]{article}

\usepackage{amsmath}
\usepackage{amsfonts}
\usepackage{amssymb}
\usepackage{graphics,graphicx,tikz}
\usepackage{soul}
\usepackage{tensor}
\usetikzlibrary{patterns}

\numberwithin{equation}{section} 
\usepackage[linktocpage]{hyperref}
\hypersetup{
	colorlinks=true,
	linkcolor=blue,
	filecolor=magenta,      
	urlcolor=blue,
	citecolor=blue
}
\def\beq{\begin{equation}}
\def\eeq{\end{equation}}
\newcommand{\commentOut}[1]{}
\def\bea{\begin{align}}
\def\eea{\end{align}}
\def\gn{\zeta}

\usepackage{color}

\begin{document}
\begin{titlepage}
\hfill \hbox{CERN-TH-2022-058}
\vskip 0.1cm
\hfill \hbox{NORDITA 2022-022}
\vskip 0.1cm
\hfill \hbox{QMUL-PH-22-14}
\vskip 0.1cm
\hfill \hbox{UUITP-20/22}
\vskip 0.5cm
\begin{flushright}
\end{flushright}
\vskip 1.0cm
\begin{center}
{\Large \bf The eikonal operator at arbitrary \\ velocities I: the soft-radiation limit}
\vskip 1.0cm {\large  Paolo Di Vecchia$^{a, b}$, Carlo Heissenberg$^{b,c}$,
Rodolfo Russo$^{d}$, \\
Gabriele Veneziano$^{e, f}$ } \\[0.7cm]

{\it \small $^a$ The Niels Bohr Institute, Blegdamsvej 17, DK-2100 Copenhagen, Denmark}\\
{\it \small $^b$ NORDITA, KTH Royal Institute of Technology and Stockholm University, \\
 Hannes Alfv{\'{e}}ns v{\"{a}}g 12, SE-11419 Stockholm, Sweden  }\\
 {\it \small $^c$ Department of Physics and Astronomy, Uppsala University,\\ Box 516, SE-75120 Uppsala, Sweden}\\
{\it \small $^d$ Centre for Theoretical Physics, Department of Physics and Astronomy,\\
Queen Mary University of London, Mile End Road, London, E1 4NS, United Kingdom.}\\
{\it \small $^e$ Theory Department, CERN, CH-1211 Geneva 23, Switzerland}\\
{\it \small $^f$ Coll\`ege de France, 11 place M. Berthelot, 75005 Paris, France}
\end{center}
\begin{abstract}
  Observables related to the real part of the gravitational eikonal, such as the deflection angle and time delay, have been found so far to have a smooth post-Minkowskian (PM) expansion whose validity extends from the non-relativistic to the most extreme ultra-relativistic (UR) regime, which smoothly connects with massless particle collisions. To describe gravitational radiation, the eikonal phase has to be promoted to a unitary operator for which we motivate a proposal and start discussing properties in the soft-radiation limit. A convergent PM expansion is found to only hold below an UR bound (discussed in the GR literature in the seventies) above which a different expansion is instead needed implying, in general, some non-analyticity in Newton's constant. In this extreme UR regime soft radiative observables receive contributions only from gravitons and  are therefore universal. This generalises the pattern discussed in~\cite{DiVecchia:2020ymx} beyond the elastic case.
\end{abstract}
\end{titlepage}

\tableofcontents

\section{Introduction}
\label{sec:intro}

In the last few years, scattering amplitude techniques have been  successfully used to compute classical observables in gravity theories for gravitational scattering and radiation  within a post-Minkowskian (PM) expansion corresponding to the usual loop expansion of quantum field theory. At 1PM and 2PM the results have been  known for some time~\cite{Bjerrum-Bohr:2018xdl,Cristofoli:2019neg,KoemansCollado:2019ggb,Cristofoli:2020uzm,Bjerrum-Bohr:2019kec} and can also be derived using the probe limit~\cite{Damour:2019lcq}. 

Three years ago, the complete conservative part of the  deflection angle, for the scattering of two scalar particles with mass $m_1$ and $m_2$, was computed in a spectacular calculation at 3PM order~\cite{Bern:2019nnu,Bern:2019crd,Kalin:2020fhe}. The result presented the puzzle that the deflection  angle was divergent at high energy in contrast  with the corresponding finite result found earlier for  the high-energy scattering of massless particles~\cite{Amati:1990xe}. 

The puzzle was eventually solved by performing a complete calculation in ${\cal{N}}=8$ massive supergravity~\cite{DiVecchia:2020ymx}, where
  the various contributions were computed in the full soft integration region rather than being restricted to the potential region~\cite{Parra-Martinez:2020dzs}. It turned out that, from the soft region, one gets extra 3PM contributions to the deflection angle physically corresponding  to  radiation-reaction effects. These precisely cancel the  high-energy divergence coming from the potential region. As a result  one gets a deflection angle that is not only  convergent 
   but  also universal at high energy. The universality follows from the intuitive fact that, at high energy, the dominant contribution comes from the massless particle with highest spin, which in theories of gravity is always the graviton. 

Immediately after, this result was extended to Einstein's gravity by two independent methods. The first one, by Damour~\cite{Damour:2020tta}, was based on the computation of the $\mathcal O(G^2)$ loss of ``angular momentum".\footnote{The definition of the angular momentum flux at null infinity is well known to suffer from a supertranslation ambiguity (see e.g.~\cite{Bonga:2018gzr}). Even its order in $G$ can vary, as recently discussed in \cite{Veneziano:2022zwh}. Damour's definition can be recovered by choosing a suitable $i \epsilon$ prescription \cite{DiVecchia:2022owy}. The results described in this paper are insensitive to this ambiguity.} When inserted in a linear response formula~\cite{Bini:2012ji},  written in  terms of the radiated angular momentum and the 1PM deflection angle, it produced the 3PM radiation-reaction part of the scattering angle.  
 The same result was obtained by us~\cite{DiVecchia:2021ndb}  by computing instead the infrared-divergent part of  the two-loop, three-particle cut, which is entirely given by the leading soft limit of the five-point amplitude, and by then using  analyticity and crossing symmetry to extract the two-loop  radiation-reaction contribution to the real part of the eikonal and thus to the deflection angle. This result is by now confirmed by extracting the classical deflection angle from  explicit two-loop calculations~\cite{Bjerrum-Bohr:2021vuf,Bjerrum-Bohr:2021din,Bjerrum-Bohr:2021wwt,Brandhuber:2021eyq}. 

The next challenge, still within the elastic case, is to go to 4PM where there exist already impressive partial results. The conservative dynamics was computed in Ref.~\cite{Bern:2021dqo,Dlapa:2021npj} and then the addition of the tail effects~\cite{Bern:2021yeh,Dlapa:2021vgp} eliminated the infrared divergence occurring in the deflection angle. The results are, however, still incomplete~\cite{Bini:2021gat,Manohar:2022dea}  because the effect of radiation has not yet been taken completely  into account.

 In this paper and a companion one \cite{tap1}  we  study classical  observables connected to the emission of radiation from the scattering of two scalar particles at arbitrary relative velocity. This problem was addressed in the seventies in pioneering papers by D'Eath \cite{DEath:1976bbo}  and by Kovacs and Thorne \cite{Kovacs:1977uw,Kovacs:1978eu}. The results in  \cite{Kovacs:1977uw,Kovacs:1978eu}, recently reproduced in \cite{Jakobsen:2021smu,Mougiakakos:2021ckm}, pointed towards a new ultra-relativistic regime where the more traditional approximations break down.
 
 In this first paper we look  at  the simplest  situation, namely the limit in which the radiated graviton's frequency is small.\footnote{Since frequency is a dimensionful quantity, this limit has to be defined in terms of a dimensionless small parameter and of an expansion in it: the relevant quantity is $\frac{\omega b}{v}$, with $b$ the impact parameter of the collision and $v$ the relative velocity of the two incoming particles.} In particular, we will study the high-energy  behaviour  of various  radiative quantities and we will show that  they also exhibit, in a very non-trivial way,  a universal behaviour at high energy.

 In the study of the elastic scattering an important quantity is the eikonal (i.e. semiclassical) approximation to the scattering amplitude, which is obtained after a resummation 
 of an infinite number of Feynman (ladder) diagrams. In particular we stress that, while the  momentum $q$ exchanged in each rung of the ladder diagram is a small quantity of $\mathcal O(\hbar)$, after the resummation  the exchanged momentum $Q$  is a large classical quantity.  The ratio between the two gives the average number of gravitons exchanged that becomes very large in the classical limit ($\hbar \rightarrow 0$).  From the eikonal phase one can then extract classical elastic observables such as the deflection angle and the Shapiro time delay.
 
 When radiation is introduced, the eikonal becomes an operator as discussed in Refs.~\cite{Amati:1990xe,Ciafaloni:2015xsr,Ciafaloni:2018uwe} in the massless case, and more generally in Ref.~\cite{Cristofoli:2021jas}. In our case 
 of  soft radiation the eikonal becomes a product of two terms: one is an operator containing the oscillators describing the graviton degrees of freedom and the other  is  the c-number eikonal describing instead the elastic degrees of freedom. This is the eikonal operator that we will use for computing classical observables such as the waveforms and the zero-frequency limit (ZFL) of the energy emission spectrum. 
 
 Since the coefficients of the harmonic oscillators of the graviton that appear in the eikonal operator depend on the momenta of the massive particles and therefore also on the momentum exchanged in the elastic process,   an important point to clarify is  what to do with it when we go to impact parameter space. The strategy  that we follow is to first interpret the exchanged momentum as the quantity $Q^\mu$, discussed above, and then, when we go to impact parameter space, to  treat it as a derivative $ -i \hbar \frac{\partial}{\partial b_\mu}$ acting only on elastic part of the eikonal.  This essentially amounts to the substitution:
 \begin{equation}
Q^\mu \rightarrow  2p \sin \frac{\Theta_s}{2}~{\hat{b}}^\mu\,,
\label{A1}
\end{equation}
 where $\Theta_s$ is the classical deflection angle, ${\hat{b}}$ is the unit vector in the direction of the impact parameter and  $p$ is the absolute value of the momentum in the centre-of-mass frame of the elastic process.  But, since the deflection angle $\Theta_s$ depends on Newton's constant $G$, we arrive at an eikonal operator that depends in a complicated way on $G$, and this feature manifests itself very clearly when we compute  classical observables such as the waveform for each polarization.
 
 In the usual PM approximation one assumes that $\Theta_s$ is a small quantity with respect to  any other kinematic variable  and then one Taylor expands the results for $\Theta_s \ll 1$. This in practice means that, in the eikonal operator,  we always neglect  terms that contain $Q^\mu$ in Eq. \eqref{A1}. The problem  however is that, when we perform this approximation on the  waveforms and keep $\Theta_s$ fixed while sending the masses of the scalar particles to zero (or equivalently while taking the high energy limit $\sigma = - \frac{p_1p_2}{m_1 m_2}\rightarrow \infty$), we obtain waveforms that diverge when the graviton is emitted along the directions of the two energetic particles.
 
 It turns out that, when we enter the region specified by
 \begin{equation}
\mbox{max}\left\{\frac{m_1}{m_2}\, \sigma\, \Theta_s^2 ,\, \frac{m_2}{m_1}\, \sigma\, \Theta_s^2 \right\}  \gtrsim 1 
\label{A2}
\end{equation}
the usual PM  expansion breaks down because we cannot neglect $Q^\mu$ any more in the eikonal operator and in this way we get well-defined waveforms  everywhere. 

Unlike the two-loop elastic eikonal, which acquires an imaginary part  from the contribution of the three-particle cut
making it non-unitary, our eikonal operator is explicitly unitary.  The reason is that its $c$-number part involves only the real part of the elastic eikonal, while the contribution of inelastic cuts comes from the normal-reordering of a unitary operator part. 

At this point we can again compute the imaginary part of the two-loop eikonal finding agreement with the results of Ref.~\cite{DiVecchia:2021ndb} provided that we are below the bound in Eq.~\eqref{A2}. However, if we take the high energy limit on this result we get a logarithmically divergent, non-universal behaviour involving the masses of the scalar particles. In order to restore universality we have to proceed as we have discussed above in the case of the waveforms, i.e.~we can no longer regard $Q^\mu$ as small compared to the masses. In this way we get a finite and universal behaviour for the contribution of the three-particle cut or equivalently of the ZFL of the radiated-energy spectrum. Our results are confirmed by the fact that we get the same high-energy limit for GR and massive ${\cal{N}}=8$ supergravity, at the price that the final result, written in terms of the deflection angle, has a non-polynomial dependence on $G$.
 In particular, for the ZFL of the radiated energy, we get the same result as in Ref.~\cite{Gruzinov:2014moa}, obtained within a classical GR approach and in Ref.\cite{Ciafaloni:2018uwe} from a scattering amplitude perspective. 

The framework we propose here relies on dressing the elastic amplitude with an operator where soft emissions exponentiate according to Weinberg's soft theorem \cite{Bloch:1937pw,Thirring:1951cz,Weinberg:1965nx,Mirbabayi:2016axw,Choi:2017ylo,Arkani-Hamed:2020gyp}. A caveat is thus in order: due to the non-linear nature of gravitational interactions there are effects, such as non-linear memory \cite{Christodoulou:1991cr,Wiseman:1991ss}, that are not captured in our approach. We leave the explicit analysis of this point from an amplitude perspective to future investigations, while remarking that instead such difficulties are absent for linear theories.
 
The paper is  organised as follows. Section \ref{sec:eikopzfl} is divided in three subsections. In the first one we recall the elastic eikonal exponentiation stressing the difference between the small momentum $q$ exchanged in a single loop order and the macroscopic momentum $Q$ exchanged after the resummation. In the second subsection we introduce the eikonal operator by dressing the elastic one with an operator suggested by the Bloch-Nordsieck/Weinberg method. In the third subsection we use the eikonal operator to compute the two waveforms in the soft limit and make contact with the linear memory effect. In Section \ref{ssec:IRd} we connect, using the eikonal operator, the infrared divergences of the background hard process, which are encoded in $\operatorname{Im}2\delta$, the number of emitted quanta and the ZFL of the spectrum of emitted energy. Section \ref{sec:aboveandbelow} is then devoted to the detailed analysis of the behaviour of explicit expressions for such quantities below and above the bound \eqref{A2}, and to highlighting how universality is restored at high energy. We conclude with a section of summary and outlook. The paper also contains two appendices. In the first one we collect various kinematic relations that we use in the paper and in the second we discuss the ZFL of the spectrum of emitted momentum.  

\newpage
\section{The Eikonal Operator in the ZFL}
\label{sec:eikopzfl}

\subsection{The elastic eikonal exponentiation}

There is by now a well tested strategy that can be used to derive the eikonal phase from $2\to 2$ quantum scattering amplitudes. Here we consider the case where the incoming and outgoing states are distinct scalars of mass $m_1$ and $m_2$ either in GR or in ${\cal N}=8$ supergravity. The leading eikonal $\delta_0$ is obtained from the tree-level in the limit where the momentum exchange $q$ is small (see App.~\ref{app:5pkin} for our conventions on the kinematics). In particular one can focus on the non-analytic terms in $q^2$ as $q^2\to 0$, which are the only source of long-range effects in impact-parameter space:
\begin{equation}
  \label{eq:delta0}
  \begin{aligned}
  {\cal A}_0(\sigma,q^2) & = \hbar \frac{8 \pi G}{q^2} \left[4 m^2_1 m^2_2 \left(\sigma^2 -\frac{\gn}{D-2}\right)\right] + \ldots\,, \\
  2 \delta_0(\sigma,b) & = \widetilde{\cal A}_0(\sigma,b) = \int \frac{d^{D-2} q}{(2\pi \hbar)^{D-2}} \frac{{\cal A}_0(\sigma,q^2)}{4 m_1 m_2\sqrt{\sigma^2-1}} e^{i \frac{bq}{\hbar}}\,,
  \end{aligned}
\end{equation}
where $\sigma$ is given in terms of the incoming momenta $\sigma = - \frac{p_1 p_2}{m_1 m_2}$ and $\gn=1$ ($\gn=0$) for GR (${\cal N}=8$ supergravity).
While the process in~\eqref{eq:delta0} involves the exchange of a single quantum, after the eikonal resummation the leading contribution to the $S$-matrix is captured by the phase $e^{2i\delta_0}$, which effectively resums infinitely many exchanges.  In the 1PM approximation, we thus have 
\begin{equation}
  \label{eq:delta0e}
  2 \delta_0(\sigma,b) = \frac{2 G \, m_1 m_2 \left(\sigma^2 -\frac{\gn}{D-2}\right)\, \Gamma\left(\frac{D-4}{2}\right) }{\hbar \sqrt{\sigma^2-1} (\pi b^2)^{\frac{D-4}{2}}}\,.  
\end{equation}
The quantities $p$ and $\sqrt{s}$ are the spatial momentum and the total initial energy in the centre-of-mass frame
\begin{equation}
  \label{eq:ps}
   p = \frac{m_1 m_2}{\sqrt{s}} \sqrt{\sigma^2-1}\,,\qquad s = m_1^2+2m_1m_2 \sigma+m_2^2\,,
   \qquad
   \sigma= \frac{1}{\sqrt{1-v^2}}\,,
 \end{equation}
with $v$ the velocity of either body as seen in the rest frame of the other one.
Via the exponentiation, the classical scattering process emerges from the exchange of a large number of soft particles as can be seen by writing the resummed leading elastic $S$-matrix element and going back to momentum space 
\begin{equation}
	\label{eq:deltaQ0}
	S^{(M)}(\sigma,Q) \simeq \int d^{D-2} b\, e^{-i \frac{bQ}{\hbar}}\, e^{2i\delta_0(\sigma,b)}.
\end{equation}
The Fourier transform above is dominated by the saddle-point approximation:\footnote{Strictly speaking, the saddle-point condition determines $b$ in terms of $Q$, and we invert this relation to work at fixed impact parameter.}
 \begin{equation}
  \label{eq:Q}
  Q^\mu_s \simeq \hbar \frac{\partial (2\delta_0)}{\partial b_\mu} \,,\qquad
  N_s \simeq \frac{|Q_s|}{|q|} \simeq \frac{4 G m_1 m_2 \, \left(\sigma^2 -\frac{\gn}{D-2}\right) \Gamma\left(\frac{D-2}{2}\right)}{\hbar \sqrt{\sigma^2-1} \pi^{\frac{D-4}{2}} b^{D-4}} \,,
\end{equation}
where $Q_s$ represents the $\hbar$-independent momentum exchanged  by virtue of the classical deflection. 
The estimate $N_s$ for the number of soft particles exchanged during the scattering is obtained by taking the total momentum transferred $|Q_s|$ and dividing it by the typical momentum of each soft particle, $q\simeq \hbar/b$ (as follows from from \eqref{eq:delta0}). Already by using the leading eikonal it is clear that $N_s$ is large and becomes infinite in the strict classical limit. The classical deflection angle $\Theta_s$ is derived  from the momentum $|Q_s|$ and at 1PM order we have
\begin{equation}
  \label{eq:Theta1PM}
  p \,\Theta_s \simeq |Q_s| \simeq \frac{4 G m_1 m_2 \, \left(\sigma^2 -\frac{\gn}{D-2}\right) \Gamma\left(\frac{D-2}{2}\right)}{\sqrt{\sigma^2-1}\, \pi^{\frac{D-4}{2}} b^{D-3}}\,.
\end{equation}

It is straightforward to formally generalise this discussion beyond the case of the 1PM elastic eikonal. One just needs to start from the long-range elastic $S$-matrix 
\begin{equation}\label{eikonalel}
	S(\sigma,b) = 1 + i\widetilde{\mathcal A}(s,b) = \left(1+2i\Delta(s,b)\right)\, e^{2i\delta(s,b)}\,,
\end{equation}
and rewrite it in momentum space
\begin{equation}
  \label{eq:deltaQ}
  S^{(M)}(\sigma,Q) = \int d^{D-2} b\, e^{-i \frac{bQ}{\hbar}}\,\left(1+2i\Delta(\sigma,b)\right) e^{2i\delta(\sigma,b)}\;.
\end{equation}
Again the classical deflection angle $\Theta_s$ is derived from the momentum $|Q_s|$ by a saddle point now related to $\delta$ instead of $\delta_0$
\begin{equation}
  \label{eq:QTh}
  Q_s^\mu = \hbar \frac{\partial \operatorname{Re}2\delta}{\partial b_\mu}\,,\qquad \sin\frac{\Theta_s}{2} = \frac{|Q_s|}{2 p}\,.
\end{equation}
In the following, for simplicity, we omit the label $s$ in $Q_s$, since we always focus on the classical saddle point.

\subsection{Bloch-Nordsieck dressing}

We would now like to include soft radiation, i.e.~the emission of real particles with very low energies, in the above picture. In the soft limit this can be done in a very efficient way by following the method of Bloch-Nordsieck~\cite{Bloch:1937pw,Thirring:1951cz} and the closely related approach by Weinberg~\cite{Weinberg:1964ew,Weinberg:1965nx} (see also very similar discussions in the context of dressed states \cite{Mirbabayi:2016axw,Choi:2017ylo,Arkani-Hamed:2020gyp}). The emission of these soft quanta exponentiates in momentum space, as we now recall focusing for the time being on graviton emissions. We will also include massless scalars and vectors that show up in $\mathcal N=8$ supergravity amplitudes at the end of this section. The $S$-matrix element for the emission of $N$ soft gravitons factorises as the matrix element $S^{(M)}$ for the background elastic process defined in \eqref{eq:deltaQ} and $N$ universal factors $f_j(k)$ expressed in terms of the polarisation $j$ of the graviton and its momentum $k$, \cite{Bloch:1937pw,Thirring:1951cz,Weinberg:1964ew,Weinberg:1965nx} 

\begin{equation}\label{eq:SMNSM}
	S^{(M)}_{s.r.,N}
	=
	\prod_{r=1}^N f_{j_r}(k_r)\,  S^{(M)}(\sigma, Q)
\,,\quad
	f_j(k) = \varepsilon^{\ast\mu\nu}_j(k)F_{\mu\nu}(k)\,,\quad
	F^{\mu\nu}(k) = \sum_{n}\frac{\kappa\, p_n^\mu p_n^\nu}{p_n\cdot k}\,,
\end{equation}
where $\kappa= \sqrt{8\pi G}$ and $n$ runs over all external states of the elastic  ``hard'' process.\footnote{As emphasized by Weinberg (see e.g. \cite{Weinberg:1995mt}), this formula only applies to the case in which the ``bare" amplitude $S^{(M)}$ is connected, hence in our case to the $i T$ part of $S = 1 + i T$. This caveat is not important for the present discussion, but assuming \eqref{eq:SMNSM} to extend to the disconnected part of the $S$-matrix (with some appropriate $i \epsilon$ regularization of the denominators) is crucial to the calculation of the angular momentum loss given in \cite{DiVecchia:2022owy}.} 
Of  course an analogous formula holds for soft absorptions, with $f_j(k)$ replaced by $-f_j^\ast(k)$. We keep graviton momenta always future-directed, while the background momenta are always regarded as outgoing (so that incoming ones are represented by past-directed vectors). For the sake of simplicity, we write explicitly only the dependence on the graviton momentum, leaving implicit the dependence on the $p_n$ which, as we will see later, will play a crucial role in our analysis.

A key step is then to introduce creation/annihilation operators for the gravitons and the other soft particles that can be produced and absorbed in the scattering process. We work with the following conventions: the canonical commutation relations are
\begin{equation}\label{key1}
		[ a^{\phantom{\dagger}}_i (k) , a_j^\dagger (k')] =\delta(\vec k,\vec k\,')\delta_{ij} \,, \qquad \delta(\vec k,\vec k') =  2 \hbar \omega (2\pi)^{D-1}  
	\delta^{D-1} 
	(\vec{k}-\vec{k}') \, ,
\end{equation}
and we define
\begin{equation}
	\label{eq:aadcom}
			\int_{\vec k} \equiv \int_{\omega<\omega_\ast} \frac{d^{D-1}\vec k}{2 \omega(2\pi)^{D-1}}\,, \qquad \omega \equiv |\vec k|\, ,
\end{equation}
where we regard $\omega$ as a frequency and $\vec k$ as a wave-vector. Following Weinberg \cite{Weinberg:1965nx}, we have also introduced a frequency scale $\omega_\ast$ below which the approximation \eqref{eq:SMNSM} is valid.\footnote{Unlike in \cite{Weinberg:1965nx}, we do not need an infrared frequency cutoff since we use dimensional regularization.}
We can then write the $S$-matrix for the emissions \eqref{eq:SMNSM} as the matrix element~\eqref{eq:deltaQ} for the elastic process dressed by an exponential factor depending on the oscillators mentioned above: 
\begin{equation}\label{}
	e^{2i\hat\delta_{s.r.}}
	=
	\exp\left(\frac1\hbar\int_{\vec k} \sum_{j} \left[ 
	f^{\phantom{\dagger}}_j(k)\, a^\dagger_j(k)
	-f^\ast_j(k)\, a^{\phantom{\ast}}_j(k)  \right]\right)
\end{equation}
in terms of which:
\begin{equation}
  \label{eq:Asr}
  S^{(M)}_{s.r.}  = 
  	e^{2i\hat\delta_{s.r.}}\,
  \frac{S^{(M)}(\sigma,Q)}{\langle 0|e^{2i\hat\delta_{s.r.}}|0\rangle} \,,
\end{equation}
where the subscript {\em s.r.} stands for soft radiation and indicates that we are restricting ourselves to $\omega < \omega_\ast$ gravitons. The denominator appearing in \eqref{eq:Asr} amounts to having taken out, momentarily, virtual graviton corrections up to the scale $\omega_\ast$, their effect being automatically reintroduced through the normal ordering of the operator in front.
After so doing, the matrix elements \eqref{eq:SMNSM} are simply  recovered using the commutation relations \eqref{key1} in
\begin{equation}\label{SMajS}
	S_{s.r.,N}^{(M)}
	=
	\langle0| a_{j_1}(k_1)\cdots a_{j_N}(k_N)S^{(M)}_{s.r.}|0\rangle\,,
\end{equation}
where the state without any graviton, $|0\rangle$, is annihilated by all $a_j$. Similarly, the quantities $\langle 0 | S_{s.r.}^{(M)} a_{j_1}^\dagger \cdots a_{j_N}^\dagger|0\rangle$ reproduce the matrix elements for soft absorptions, with the appropriate signs.

To leading order in the soft expansion, it is possible to work with~\eqref{eq:Asr} by using the momenta of energetic external particles in the elastic process as given and this is the viewpoint taken in~\cite{Weinberg:1964ew,Weinberg:1965nx} and more recently in~\cite{Laddha:2018vbn,Sahoo:2018lxl,Saha:2019tub,Addazi:2019mjh,Sahoo:2021ctw}. Here we would like to include the dynamical information about the scattering process, specifying that in our case it is due to the gravitational interaction itself. This is more easily done by going back to impact parameter space as the eikonal takes the compact expression~\eqref{eikonalel} for the elastic process. 

We first consider the Fourier transform to $b$-space of the two factors in~\eqref{eq:Asr} separately. By construction the second factor, which describes the elastic process, gives the eikonal~\eqref{eikonalel}. However, thanks to the division by $\langle 0|e^{2i\hat\delta_{s.r.}}|0\rangle$ in \eqref{eq:Asr}, one needs only the real part of $\delta$, as the divergent imaginary part is automatically encoded in the new operator part,\footnote{Even after having taken out soft real and virtual radiation, the true $S$-matrix connects the two particle initial state to other inelastic channels (e.g.~via emission of gravitons with $\omega > \omega_\ast$). As a consequence, some finite imaginary part will be left over in $\delta$.} as we shall see in Section~\ref{ssec:IRd}. The first factor in~\eqref{eq:Asr} is instead regular as $Q\to 0$, so we can write it as a differential operator acting on a delta-function $\delta^{D-2}(b)$ trading each $Q$ with a derivative
\begin{equation}
  \label{eq:qtodn}
   Q^\mu \to  -i \hbar \frac{\partial}{\partial b_\mu}
\end{equation}
in the Fourier transform.
Of course the product of these two factors in~\eqref{eq:Asr} becomes a convolution in $b$ space. However, since one factor is just a delta function, the integral of the convolution can be performed straightforwardly, and one obtains
\begin{equation}
  \label{eq:eiksr}
  \begin{aligned}
   S_{s.r.}(\sigma,b;a,a^\dagger)  =\ & \exp\left(\frac1\hbar\int_{\vec k} \sum_{j} \left[f_j(k) a^\dagger_j(k)- f^\ast_j(k) a_j(k) \right]
   \right) \\
   & \left[1+2i\Delta(\sigma,b)\right] e^{i \operatorname{Re}2\delta(\sigma,b)} \,,
 \end{aligned}
\end{equation}
where the external momenta $p_n$ in the first line contain derivatives acting on the $b$-dependent functions in the second line.  In the ZFL we can use the 4-point kinematics and identify the momentum transferred in Eqs.~\eqref{eq:inpie} and~\eqref{eq:inpfe} with a derivative acting on the elastic eikonal as in~\eqref{eq:qtodn}. Since the soft operator becomes proportional to $Q$ we see explicitly how the disconnected term of the elastic scattering amplitude drops out.

Let us make two general comments before using~\eqref{eq:eiksr} in some concrete calculations. First, the classical $S$-matrix obtained by neglecting the quantum remainder $\Delta$ is explicitly unitary since  only the real part of the elastic $\delta$ enters in this equation and the inelastic prefactor is the exponential of an anti-Hermitian operator.\footnote{Actually, the situation is not so simple. Although each individual graviton carries a negligible amount of energy ${\cal O}(\frac{\hbar}{b})$, the total amount of radiated energy is a classical quantity to be compared with the other classical energies in the problem. We will discuss elsewhere how explicit energy conservation can be added while pushing unitarity violations to higher orders in the PM expansion.}

Second, when focusing on the classical part, the derivatives hidden in the external momenta $p_n$ due to \eqref{eq:qtodn} can be always taken to act  on the eikonal phase:
for any smooth function $\varphi(Q)=\sum_nc_n (Q^2)^n$, performing the replacement \eqref{eq:qtodn}, 
\begin{equation}\label{scalingargument}
\begin{split}
		\varphi(Q)\, e^{i\operatorname{Re}2\delta}\to\sum_nc_n (-i\hbar\partial_b)^{2n} e^{i\operatorname{Re}2\delta}
	&= \sum_n c_n \left[\hbar
	\frac{\partial\operatorname{Re}2\delta}{\partial b}
	\right]^{2n} e^{i\operatorname{Re}2\delta}
	+
	\mathcal O(\hbar)\\
	&=
	\varphi\left(
	\hbar\,
		\frac{\partial\operatorname{Re}2\delta}{\partial b}\right) e^{i\operatorname{Re}2\delta} +\mathcal O(\hbar)\,,
\end{split}
\end{equation}
i.e.
\begin{equation}
	\label{qtoQ2}
	Q^\mu \to \hbar\, \frac{\partial \operatorname{Re}2\delta}{\partial b_\mu} = \hat{b}^\mu\, 2 p \sin\frac{\Theta_s}{2}\,,
\end{equation}
where $\hat{b}^\mu = b^\mu/|b|$.
Indeed, $\hbar \partial_b \operatorname{Re}2\delta\sim\mathcal O(\hbar^0)$, while if we were act on $\operatorname{Re}2\delta$ more than once with $\hbar\partial_b$, we would only produce terms of higher order in $\hbar$.

Then we effectively should use the following momenta in~\eqref{eq:eiksr} for the external hard particles 
\begin{gather}
  \label{eq:ppartb}
  p_1^\mu = -{\overline{m}}_1 u_1^\mu + \hat{b}^\mu\,  p \sin\frac{\Theta_s}{2}\;,\quad p_2^\mu = -{\overline{m}}_2 u_2^\mu - \hat{b}^\mu\,  p \sin\frac{\Theta_s}{2}\;,\\ \nonumber
  p_4^\mu= {\overline{m}}_1u_1^\mu + \hat{b}^\mu\,  p \sin\frac{\Theta_s}{2} \;,\quad
  p_3^\mu={\overline{m}}_2u_2^\mu - \hat{b}^\mu\,  p \sin\frac{\Theta_s}{2} \;,
\end{gather}
which are simply the initial and the final momenta in the classical elastic scattering. In a PM expansion it seems that one can discard the terms involving $\sin\frac{\Theta_s}{2}$ since $\Theta_s$ is proportional to $G$, see~\eqref{eq:delta0e}. This is equivalent to expanding for small $Q$ the first line of~\eqref{eq:eiksr} as done in~\cite{DiVecchia:2021ndb}, see the step between~(2.9) and~(2.11) of that reference. However this expansion is not justified in all kinematic regimes, as we will see below in the discussion of the waveforms in the ZFL.

In the following, we will apply the eikonal operator to discuss the contribution of low-energy gravitons to several observables, including the waveforms, memory, and the particle-energy emission spectrum. The general strategy, given any quantum observable $\mathcal O$, is to take its expectation value according to
\begin{equation}\label{SOS}
	\langle \mathcal O \rangle = \langle 0| S_{s.r.}^\dagger \mathcal O \, S_{s.r.} |0\rangle\,.  
\end{equation}
Physically, this means to evaluate the mean value of $\mathcal O$ in the final state of the scattering event, obtained by applying $S_{s.r.}$ to the state with no gravitons.
It can also be instructive to inspect more closely the dependence of this classical value on the number of exchanged gravitons (or other massless particles), and for this reason it is useful to insert a complete set of Fock states, so that
\begin{equation}\label{eq:O}
\langle \mathcal O \rangle = \sum_{N=0}^\infty \langle \mathcal O \rangle_N\,,
\end{equation}
with
\begin{equation}\label{eq:ON}
	\langle \mathcal O \rangle_N 
	=
	 \frac{1}{N!}\,
	\sum_{j_1,\ldots,j_N} \int_{\vec k_1}\cdots \int_{\vec k_N}\,
 	\langle 0|S^\dagger_{s.r.} \mathcal O a_{j_1}^\dagger\cdots a_{j_N}^\dagger |0\rangle
 	\langle 0|a_{j_1}\cdots a_{j_N} S_{s.r.} |0\rangle\,.
\end{equation}

We conclude this section by describing how the soft eikonal operator is modified to include the presence of other massless fields (scalars and vectors), which will be useful to discuss the case of $\mathcal{N}=8$ supergravity where the massive particles are described by KK modes. As mentioned in the introduction, this is a nice toy model which has the same features of GR but provides simpler results. The $S$-matrix elements for soft emissions factorize in a way analogous to \eqref{eq:SMNSM}, with soft factors that instead of $f_j(k)$ are given by
\begin{equation}\label{fvecfsc}
	f_j^\text{vec}(k)=\sum_{n} \eta_n e_n \frac{\varepsilon^\ast_{\mu,j}(k) p_n^\mu}{p_n\cdot k},\qquad
	f^\text{sc}_j(k) = \sum_{n} \frac{g_n}{p_n\cdot k}
\end{equation}
for vectors and scalars respectively; here $\eta_n$ takes the value $+1$ for outgoing and $-1$ for incoming states while $e_n$ and $g_n$ denote suitable couplings.
These new soft particles are easily accommodated in the eikonal operator: it is sufficient to include in the first line of~\eqref{eq:eiksr} the relevant operators $a_d$ for the dilaton (with coupling $g_n = -\kappa\,m_n^2/\sqrt{D-2}$)  and $a_{v_i,j}$ for two vectors ($e_n=\sqrt 2\,\kappa\, m_i$) and $a_{s_i}$ for two scalars ($g_n = \kappa\, m_i^2$). Such vectors and scalars arise in the KK-compactification and couple to particles of mass\footnote{A general way to introduce masses in the $\mathcal{N}=8$ context is discussed in~\cite{Caron-Huot:2018ape}; here we work in the particular case discussed in~\cite{Parra-Martinez:2020dzs} and further focus on the case $\sin\phi=1$ in their notation, see also~\cite{DiVecchia:2021ndb}.} $m_i$ ($i=1,2$). Thus the $\mathcal{N}=8$ eikonal operator takes the same form as in the GR case
\begin{equation}
  \label{eq:d}
   S^{\mathcal{N}=8}_{s.r.}(\sigma,b;a,a^\dagger) = e^{2i\hat{\delta}^{\mathcal{N}=8}_{s.r.}} \left[1+2i\Delta_{\mathcal{N}=8}(\sigma,b)\right] e^{2i\operatorname{Re}\delta^{\mathcal{N}=8}(\sigma,b)}\;,
\end{equation}
where
\begin{align}
    2i\hat{\delta}^{\mathcal{N}=8}_{s.r.} & =\frac{1}{\hbar}\int_{\vec k} \sum_{j} 
    \left[ (f_j a^\dagger_j-f^\ast_j a_j) + (f^{d} a_d^\dagger- f^{d\ast}a_d) +  \right. \nonumber \\ & ~ + \left. (f^{v}_{j} a^\dagger_{v,j}-f^{v\ast}_{j} a_{v,j}) + (f^{s}_{j} a^\dagger_{s,j}-f^{s\ast}_{j} a_{s,j}) \right],   \label{eq:dn8sr}   
\end{align}
where, as before, $j$ labels the various physical polarisations, $\eta_n$ takes the value $+1$ for $n=3,4$ and $-1$ for $n=1,2$, and the massive particles are pairwise identical ($m_3=m_2$, $m_4=m_1$) and so they couple to the same soft KK modes ($a_{s_3} = a_{s_2}$, $a_{s_4} = a_{s_1}$, etc.); finally we refer to~\cite{DiVecchia:2021bdo} for the first 3PM order of the elastic data $\Delta^{\mathcal{N}=8}$ and $\delta^{\mathcal{N}=8}$.

\subsection{Soft waveforms and memory}
\label{ssec:wavzfl}

As a first application of the above tools, let us discuss how one obtains the leading-order as $\omega\to0$ of the asymptotic waveforms.
The classical field is obtained by inserting in the expectation value \eqref{SOS} the free gravitational field 
\begin{equation}\label{}
	H_{\mu\nu}(x)=
	\int_{\vec k} \sum_j \left[ \varepsilon_{j,\mu\nu}(k) a_j(k)\, e^{ikx} + \varepsilon^\ast_{j,\mu\nu}(k) a^\dagger_j(k)\,\,e^{-ikx}\right],
	\qquad
	\langle H_{\mu\nu}(x) \rangle = h_{\mu\nu}(x)\,.
\end{equation}
This yields,
\begin{equation}
	\label{eq:hmnzfl}
	h_{\mu\nu}(x)=\int_{\vec k} \sum_j  \,\langle 0| S_{s.r.}^\dagger  \left[ \varepsilon_{j,\mu\nu}(k) a_j(k)\, e^{ikx} + \varepsilon^\ast_{j,\mu\nu}(k) a^\dagger_j(k)\,\,e^{-ikx}\right] S_{s.r.}|0\rangle\,.
\end{equation}
The oscillators inserted in the expectation value act in a straightforward way on the first line of the $S$-matrix in~\eqref{eq:eiksr}, so that the operator part of the $S$-matrix cancels:
\begin{equation}
	\label{eq:hmnzfl2}
	h_{\mu\nu}(x)= \int_{\vec k}   e^{-2i\operatorname{Re}\delta}    \left[ f_{\mu\nu}(k)\, e^{ikx} + f_{\mu\nu}^\ast(k)\,e^{-ikx}\right] e^{ 2i\operatorname{Re}\delta}\,.
\end{equation}
In this equation,
\begin{equation}\label{fTT}
	f^{\mu\nu}(k) = \Pi^{\mu\nu}_{\rho\sigma}(\hat k) F^{\rho\sigma}(k)\,,
	\qquad
	F^{\mu\nu}(k)
	=
	\sum_n \frac{\kappa\,  p_n^\mu p_n^\nu}{p_n\cdot k}\,,
\end{equation}
and $\Pi^{\mu\nu}$ is the usual transverse-traceless projector over physical degrees of freedom.
Letting derivatives with respect to $b$ act as in \eqref{qtoQ2}, we can simply write
\begin{equation}
	\label{eq:hmnzfz}
	h_{\mu\nu}(x) = \int_{\vec k} \left[    f_{\mu\nu}(k)  \,e^{ikx} +   f^\ast_{\mu\nu}(k) \,e^{-ikx}\right],
\end{equation}
keeping in mind that now the external momenta should be treated as in \eqref{eq:ppartb}.

Now we consider the asymptotic limit for the gravitational field, where $x^\mu = (x^0,\vec x) = (u+r,r\hat x)$ and the detector's distance is taken to infinity, $r\to\infty$, for fixed retarded time $u$ and angles $\hat x$.
In this limit, a standard stationary-phase argument (see e.g.~\cite{Strominger:2014pwa,Campoleoni:2019ptc,Campoleoni:2020ejn}) yields
\begin{equation}\label{}
	h_{\mu\nu} (u+r,r \hat x) \sim \frac{1}{2(2i\pi r)^{\frac{D-2}{2}}}
	\int_0^\infty \frac{d\omega}{2\pi}\,  \omega^{\frac{D-4}{2}} \,
	f_{\mu\nu}(\omega,\omega \hat x)\, e^{-i\omega u} 
	+
	{c.c.}\,,
\end{equation}
where $c.c.$ stands for the complex conjugate of the term explicitly written.
Note that in this step the angular integral over the momenta $\hat k$ localizes along the observation direction $\hat x$.
Focusing on the four-dimensional case and recalling $f_{\mu\nu}(k)=-f_{\mu\nu}^\ast(-k)$ leads to
\begin{equation}\label{hmunufmunu}
	h_{\mu\nu} \sim 
	\frac{1}{4\pi r}
	\int_{-\infty}^{+\infty} \frac{d\omega}{2i\pi}\, 
	f_{\mu\nu}(\omega,\omega \hat x)\, e^{-i\omega u} \,.
\end{equation}
where the two terms in Eq.~\eqref{eq:hmnzfz} recombined to reconstruct a single integral over positive and negative frequencies \cite{Cristofoli:2021vyo}.
Adjusting the overall normalization by comparing
\begin{equation}\label{}
	g_{\mu\nu} = \eta_{\mu\nu} + 2 W_{\mu\nu}  = \eta_{\mu\nu} + 2\kappa h_{\mu\nu}\,,
\end{equation}
we also define 
\begin{equation}\label{}
	W_{\mu\nu} =  \kappa\, h_{\mu\nu} \,.
\end{equation}
Performing the Fourier transform in \eqref{hmunufmunu}  requires in principle to specify how the $1/\omega$ singularity at $\omega=0$ is circumvented  \cite{Laddha:2018vbn,Sahoo:2018lxl,Saha:2019tub,Sahoo:2021ctw,DiVecchia:2022owy}. However, as stressed in \cite{Strominger:2014pwa,Strominger:2017zoo}, the key point is that the behaviour of the waveform at large $|u|$ is completely determined by this pole at $\omega=0$, and possible ambiguities are in fact $u$-independent. 
Considering the invariant combination
\begin{equation}\label{}
	\Delta W_{\mu\nu}(\hat x) = W_{\mu\nu}(u>0,\hat x)-W_{\mu\nu}(u<0,\hat x)\,,
\end{equation}
we thus obtain
\begin{equation}\label{}
	\Delta W_{\mu\nu}(\hat x)
	=\frac{2G}{r}\,\Pi^{\mu\nu}_{\rho\sigma}(\hat x)\,\sum_{n} \frac{\eta_n\, p_n^\rho p_n^\sigma}{E_n-\vec k_n\cdot\hat x}\,,
\end{equation}
where $p_n = \eta_n (E_n, \vec k_n)$.
In this way, we reproduce the well-known memory effect \cite{Zeldovich:1974gvh} i.e.~the leading result of~\cite{Laddha:2018vbn,Sahoo:2018lxl,Saha:2019tub,Sahoo:2021ctw} or the term indicated as $A_{\mu\nu}$ in~\cite{Sahoo:2021ctw}.
We refer to Ref.~\cite{DiVecchia:2022owy} for further discussion of the evaluation of the waveform in $u$-space, in particular with the Feynman $-i0$ prescription.
Moreover, let us recall that our approach does not capture non-linear memory effects \cite{Christodoulou:1991cr,Wiseman:1991ss,Damour:2020tta}. 

For the remainder of this section, let us refrain from performing this Fourier transform and remain in $\omega$-space, defining
\begin{equation}\label{}
	W^{LS}_{\mu\nu}  = \frac{\kappa}{4\pi r}\, F_{\mu\nu}
\end{equation}
with $F^{\mu\nu}$ as in \eqref{fTT}, up to the identification $\hat k =\hat x$.
Let us also specialize our expressions using the following kinematics
\begin{equation}
	\begin{split}
		p_1 &=  \left(-E_1,{\hat{b}}\,p \sin \frac{\Theta_s}{2}, -  p \cos \frac{\Theta_s}{2}\right)\!,\quad
		p_2 =  \left(-E_2,-{\hat{b}}\, p\sin \frac{\Theta_s}{2}, + p \cos \frac{\Theta_s}{2}\right)\!, 
		\\ 
		p_4 &= \left(+E_1, {\hat{b}} \, p\sin \frac{\Theta_s}{2}, + p \cos \frac{\Theta_s}{2}\right)\!,\quad
		p_3 = \left(+E_2, -{\hat{b}}\,p \sin \frac{\Theta_s}{2}, -p\cos \frac{\Theta_s}{2}\right)\!,
	\end{split}
	\label{WF12}
\end{equation}
which follows from~\eqref{eq:ppartb}, and consider the projection on the two polarizations described in Appendix~\ref{app:5pkin} (see Eqs.~\eqref{GGG8}, \eqref{eq:timesplus})
\begin{equation}\label{key}
	W_{\times}^{LS} = \frac{\kappa}{4\pi r}\, \epsilon_{\phi}^\mu\epsilon_{\theta}^\nu\,F_{\mu\nu}\,,\qquad
	W_{+}^{LS} = \frac{\kappa}{4\pi r}\,\frac{1}{2}\left( \epsilon_{\theta}^\mu\epsilon_{\theta}^\nu - \epsilon_{\phi}^\mu\epsilon_{\phi}^\nu\right) F_{\mu\nu}\,.
\end{equation}
We obtain for the leading soft term of the $\times$ polarization
\begin{align}
	\label{eq:wavAet}
	& W^{LS}_{\times}(k) = \frac{2 G}{r} \frac{p}{\omega} \left(-\sin\phi \sin\frac{\Theta_s}{2}\right) \times \\ \nonumber & \left[ \frac{  \cos \theta \cos \phi \sin \frac{\Theta_s}{2} + \sin \theta \cos \frac{\Theta_s}{2} }{ \frac{E_1}{p} + \sin \frac{\Theta_s}{2} \sin \theta \cos \phi- \cos \frac{\Theta_s}{2} \cos \theta} + \frac{  \cos \theta \cos \phi \sin \frac{\Theta_s}{2} + \sin \theta \cos \frac{\Theta_s}{2} }{ \frac{E_2}{p} -( \sin \frac{\Theta_s}{2} \sin \theta \cos \phi- \cos \frac{\Theta_s}{2} \cos \theta)}\right.\\ \nonumber & \left. - \frac{ \cos \theta \cos \phi \sin \frac{\Theta_s}{2} - \sin \theta \cos \frac{\Theta_s}{2} }{\frac{E_1}{p} - \sin \frac{\Theta_s}{2}  \sin \theta \cos \phi -  \cos \frac{\Theta_s}{2}  \cos \theta} - \frac{ \cos \theta \cos \phi \sin \frac{\Theta_s}{2} - \sin \theta \cos \frac{\Theta_s}{2} }{\frac{E_2}{p} + \sin \frac{\Theta_s}{2}  \sin \theta \cos \phi + \cos \frac{\Theta_s}{2}  \cos \theta} \right],
\end{align}
where we recall that $\Theta_s$ stands for its classical expression in terms of $b$ and we aligned $\hat b$ along the $x$-axis.
Similarly for the other polarisation we have

\begin{equation}\label{}
	\begin{split} 
		W^{LS}_+   = \frac{G}{r} \frac{p}{\omega} 
		& \Bigg[ \frac{\left(\sin \frac{\Theta_s}{2}\cos \phi \cos \theta + \sin \theta  \cos\frac{\Theta_s}{2}\ \right)^2-\sin^2\phi \sin^2 \frac{\Theta_s}{2}}{\frac{E_1}{p} + \sin \frac{\Theta_s}{2} \sin \theta \cos \phi - \cos\frac{\Theta_s}{2}\cos \theta} \\
		& + \frac{\left(\sin \frac{\Theta_s}{2}\cos \phi \cos \theta + \sin \theta  \cos\frac{\Theta_s}{2}\ \right)^2-\sin^2\phi \sin^2 \frac{\Theta_s}{2}}{\frac{E_2}{p} - \sin \frac{\Theta_s}{2} \sin \theta \cos \phi + \cos\frac{\Theta_s}{2}\cos \theta}  \label{WF43} \\
		&+ \frac{\left(\sin \frac{\Theta_s}{2}\cos \phi \cos \theta - \sin \theta  \cos\frac{\Theta_s}{2}\ \right)^2-\sin^2\phi \sin^2 \frac{\Theta_s}{2}}{-\frac{E_1}{p} + \sin \frac{\Theta_s}{2} \sin \theta \cos \phi + \cos\frac{\Theta_s}{2}\cos \theta} \\ 
		& + \frac{\left(-\sin \frac{\Theta_s}{2}\cos \phi \cos \theta + \sin \theta  \cos\frac{\Theta_s}{2}\ \right)^2-\sin^2\phi \sin^2 \frac{\Theta_s}{2}}{-\frac{E_2}{p} - \sin \frac{\Theta_s}{2} \sin \theta \cos \phi - \cos\frac{\Theta_s}{2}\cos \theta} \Bigg].
	\end{split}
\end{equation}

The results~\eqref{eq:wavAet} and~\eqref{WF43} have a complicated dependence on Newton's constant through $\Theta_s$ and so on Newton's constant $G$. The usual way of performing the PM expansion on the waveforms is to assume that $\Theta_s$ is small with respect to any other kinematic ratio and then Taylor expand the results for $\Theta_s\ll 1$. At the leading PM order we have
	\begin{align}
		W^{LS}_{\times}(k) & \simeq - \frac{2 G}{r} \frac{p \,\Theta_s}{\omega} \sin\phi \sin \theta  \left[ \frac{1}{ \frac{E_1}{p} - \cos \theta} + \frac{1}{ \frac{E_2}{p} + \cos \theta}\right],   \label{eq:wavAetPM1}\\
		W^{LS}_{+}(k) & \simeq \frac{2G}{r} \frac{\Theta_s}{\omega}  \cos \phi \sin \theta \Bigg\{ -\frac{p}{2} \sin^2 \theta   \label{eq:wavAepPM1} \\
		& \times \Bigg[ \frac{1}{(\frac{E_1}{p} - \cos \theta)^2} - \frac{1}{(\frac{E_2}{p} +\cos \theta)^2} \Bigg] + \frac{ \cos \theta \sqrt{s}}{ (\frac{E_1}{p} - \cos \theta)(\frac{E_2}{p} + \cos \theta)} \Bigg\},
		\nonumber
	\end{align}
which agree with Ref.~\cite{Damour:2020tta} and Eq.~(4.5) of~\cite{DiVecchia:2021ndb}. Of course the full leading PM waveforms~\cite{Kovacs:1978eu,Jakobsen:2021smu,Mougiakakos:2021ckm,Riva:2021vnj} reduce to the above result  in the ZFL.

Notice however that~\eqref{eq:wavAetPM1}, \eqref{eq:wavAepPM1} diverge when $\Theta_s$ is small but fixed, we send $m_i$ to zero (or equivalently we take $\sigma \to \infty$) and we also take the $\theta \to 0$ limit. For $W^{LS}_{\times}$ this is already visible at the leading ${\cal O}(\Theta_s)$ term: in the limit $E_i/p \to 1$, the denominators vanish quadratically as $\theta \to 0$ or $\theta \to\pi$ while the prefactor vanishes just linearly. However, when $\theta\simeq \Theta_s \ll 1$ the corrections to the expressions in the denominators in~\eqref{eq:wavAetPM1} become important and in particular they compete with the term $E_i/p - 1$ when 
\begin{equation}
	\label{eq:KTb}
	\mbox{max}\left\{\frac{m_1}{m_2}\, \sigma\, \Theta_s^2 ,\, \frac{m_2}{m_1}\, \sigma\, \Theta_s^2 \right\}  \gtrsim 1\,,\quad
	\text{i.e.}\quad \mbox{max}\left\{\frac{Gs}{m_2 b}  ,\, \frac{Gs}{m_1 b} \right\}  \gtrsim 1\,,	
\end{equation}
where, in the latter formulation, we have rewritten $\Theta_s$ in terms of $b$.
When entering the kinematic region defined above, the usual PM expansion which yields~\eqref{eq:wavAetPM1} and~\eqref{eq:wavAepPM1} breaks down. In this region one needs to follow the approach discussed from~\eqref{eq:qtodn} in order to treat properly the collinear radiation and, as we will see, this will be important also for the energy spectrum. This phenomenon was discussed long ago from the GR perspective, see\footnote{While we here refer to the deflection angle $\Theta_s$ in the centre-of-mass frame, the one quoted in the abstract of~\cite{Kovacs:1978eu} should be the angle $\Theta_s'$ in frame where the heavier particle (say 1) is initially at rest. They are related by $\tan{\Theta_s'}=m_1\sin\Theta_s/(E_2+E_1\cos\Theta_s)$, so that for very large velocities $\sigma\gg1$ one gets $\tan\Theta_s' \simeq \sqrt{2 m_1/(m_2\sigma)}\,\sin\Theta_s/(1+\cos\Theta_s) $. To first order in the PM expansion we thus have $\Theta_s' \simeq \sqrt{m_1/(2m_2\sigma)} \Theta_s$ and so~\eqref{eq:KTb} reads $\sigma\,\Theta_s' \gtrsim 1$.}~\cite{Kovacs:1978eu} and~\cite{Damour:2019lcq} which refer to~\cite{DEath:1976bbo}.

In the massless limit we have $E_i/p\to 1$, $p\to \sqrt{s}/2$ and the waveforms simplify to
\begin{equation}
	\label{eq:wavAetm0}
	W^{LS}_{\times}(k) = -\frac{2 G}{r} \frac{\sqrt{s}}{\omega} \sin\phi \sin\frac{\Theta_s}{2} \sum_{\alpha=\pm 1} \left[ \frac{ \alpha \cos \theta \cos \phi \sin \frac{\Theta_s}{2} +\sin \theta \cos \frac{\Theta_s}{2} }{1 -( \sin \frac{\Theta_s}{2} \sin \theta \cos \phi-\alpha \cos \frac{\Theta_s}{2} \cos \theta)^2}\right]
\end{equation}
and
\begin{equation}
	\label{eq:wavAepm0}
	W^{LS}_{+}(k) = -\frac{2 G}{r} \frac{\sqrt{s}}{\omega} \sum_{\alpha=\pm 1} \left[ \frac{\alpha \sin^2\phi \sin^2\frac{\Theta_s}{2} }{1 -( \sin \frac{\Theta_s}{2} \sin \theta \cos \phi-\alpha \cos \frac{\Theta_s}{2} \cos \theta)^2}\right].
\end{equation}
As mentioned above, the dominant PM contributions to the ZFL come from the polarisation $W^{LS}_{\times}$ and is of order $\Theta_s$, while $W^{LS}_{+}$ is of order $\Theta^3_s$.

We conclude this section by mentioning that an entirely analogous discussion holds for the corresponding waveforms associated to massless vector and scalar radiation, which are relevant in particular for the $\mathcal N=8$ setup. As an example let us briefly focus on the case of the dilaton emission. From~\eqref{fvecfsc} with $g_n = -\kappa m_n^2/\sqrt{D-2}$ we have the following soft waveform
  \begin{equation}
    \label{eq:Wdil}
    \begin{aligned}
     W^{LS}_{\rm dil}(k) = -\frac{2 G}{\sqrt{D-2}\, r} \frac{1}{\omega} & \sum_{\alpha=\pm 1} \left[ \frac{ \alpha m_1^2 }{ E_1 + \alpha p\sin \frac{\Theta_s}{2} \sin \theta \cos \phi-p \cos \frac{\Theta_s}{2} \cos \theta}\right. \\ & +\left. \frac{\alpha m_2^2 }{ E_2 - \alpha p \sin \frac{\Theta_s}{2} \sin \theta \cos \phi + p \cos \frac{\Theta_s}{2} \cos \theta}\right].
   \end{aligned}
 \end{equation}
At leading PM order below the threshold~\eqref{eq:KTb}, there is agreement with the ZFL of the results in~\cite{DiVecchia:2021bdo}
\begin{equation}
  \label{eq:Wdils}
  W^{LS}_{\rm dil}(k) = \frac{2 G}{\sqrt{D-2}\, r} \frac{p\Theta_s}{\omega} \sin\theta \cos\phi \left[\frac{m_1^2}{(E_1-p\cos\theta)^2}-\frac{m_2^2}{(E_2+p\cos\theta)^2}\right].
\end{equation}
As for the graviton case, also~\eqref{eq:Wdils} has an anomalous behaviour when $\sigma\to \infty$ and $\sin\theta \sim\sqrt{2/\sigma}$, {\rm i.e.}~the square parenthesis scales as $\pm2 \nu \sigma$ (where $\nu = m_1 m_2/(m_1+m_2)^2$) and so it becomes large. This means that the dilaton yields non trivial contributions for large velocities but below the threshold~\eqref{eq:KTb}. However, above that bound, the approximations yielding~\eqref{eq:Wdils} break down and one has to use~\eqref{eq:Wdil}, which vanishes in the extreme UR regime. So, as expected, the dilaton decouples at very high energies.

\section{Infrared Divergences and Soft Spectra}
\label{ssec:IRd}

Using the eikonal operator we can show explicitly that the amplitude, and thus the transition probability, for the purely elastic $2\to 2$ process is exponentially suppressed, in fact infinitely so in $D=4$. As we shall review below, this infrared divergence is intimately connected to the fact that an infinite number of field quanta is emitted via classical radiation. Moreover, it is also directly related to a (finite) observable quantity: the zero-frequency limit of the spectrum of emitted energy. 

\subsection{Infrared divergences in \texorpdfstring{$\operatorname{Im}2\delta(\sigma,b)$}{Im2d}}

Let us take the final state $S_{s.r.}|0\rangle$ and project it on the graviton vacuum $|0\rangle$. Then one needs to normal order the inelastic exponential through the usual Baker--Campbell--Hausdorff formula
\begin{equation}
e^{v a^\dagger - v^* a}= e^{va^\dagger} e^{-v^*a} e^{-\frac{1}{2}|v|^2{[a,a^\dagger]}}\,,
\label{BCH}
\end{equation}
so the amplitude for the elastic process is given by
\begin{equation}\label{FPIF}
	 \langle 0| S_{s.r.}|0\rangle = \exp\left[-\frac{1}{2\hbar}\int_{\vec k}F^\ast_{\mu\nu}(k)\Pi^{\mu\nu,\rho\sigma}F_{\rho\sigma}(k)\right] e^{i\operatorname{Re}2\delta(\sigma,b)} \,,
\end{equation}
where $\Pi^{\mu\nu,\rho\sigma}$ is the usual transverse-traceless projector and we focused on the classical contribution (ignoring $\Delta$).
The transversality condition $k^\mu F_{\mu\nu}=0$, which holds for gravity by momentum conservation, grants
\begin{equation}
	\label{cancellation}
	 F^{\ast}_{\mu\nu}(k)\Pi^{\mu\nu,\rho\sigma}(k)F_{\rho\sigma}(k) = F^{\ast}_{\mu\nu}(k) P^{\mu\nu,\rho\sigma}F_{\rho\sigma}(k)\,,
\end{equation}
where
\begin{equation}\label{}
	P^{\mu\nu,\rho\sigma} = \frac12\left( \eta^{\mu\rho} \eta^{\nu\sigma} + \eta^{\mu\sigma} \eta^{\nu\rho}  -\frac{2}{D-2}\, \eta^{\mu\nu}\eta^{\rho\sigma} \right).
\end{equation}
The first exponential in~\eqref{FPIF} is a damping factor that can be interpreted as an imaginary contribution to the classical eikonal: in this way we have
\begin{equation}\label{expi2delta}
	\langle 0| S_{s.r.}|0\rangle =  e^{2i \delta(\sigma,b)}\,,
\end{equation}
where $2\delta= \operatorname{Re}2\delta + i \operatorname{Im}2\delta$ and
\begin{equation}
	\label{eq:Imdeltadef}
	\operatorname{Im}2\delta(\sigma,b) = \frac{1}{2\hbar}\int_{\vec k}F^\ast_{\mu\nu}(k) P^{\mu\nu,\rho\sigma}F_{\rho\sigma}(k)\,.
\end{equation}
This is the analogue, in $b$-space, of the damping due to summing the contributions of virtual gravitons to the elastic amplitude \cite{Weinberg:1965nx}
(we recall that the dependence on the impact parameter $b$ is implicit through the identification~\eqref{eq:qtodn}). 
Following these steps, one can also rewrite the eikonal operator directly as
\begin{equation}
	\label{BN1}
	S_{s.r.}(\sigma,b;a,a^\dagger)  = 
	\exp\left(\frac1\hbar\int_{\vec k}\sum_{j} f_j a^\dagger_i  \right)  
	\exp\left(-\frac1\hbar \int_{\vec k}\sum_{j} f^\ast_j a_j    \right) e^{2i\delta(\sigma,b)}\,.
\end{equation}

For later convenience, before evaluating it explicitly, let us write the integral in \eqref{eq:Imdeltadef} introducing the integration over the frequency with an upper cutoff $\omega_\ast$ and over the angles, using $k^\mu = \omega(1,\hat x)$ and $F^{\mu\nu}(k) = \omega^{-1}F^{\mu\nu}(1,\hat x)$,
\begin{equation}
	\label{eq:Imdeltaomegaangles}
	\operatorname{Im}2\delta(\sigma,b) = \frac{1}{2\hbar} \int_0^{\omega_\ast}\frac{d\omega}{\omega^{5-D}}\oint F^\ast_{\mu\nu}(1,\hat x) P^{\mu\nu,\rho\sigma}F_{\rho\sigma}(1,\hat x)\,\frac{d\Omega(\hat x)}{2(2\pi)^{D-1}}\,,
\end{equation}
so that
\begin{equation}
	\label{eq:Imdeltaomegaanglesint}
	\operatorname{Im}2\delta(\sigma,b) = \frac{1}{2\hbar} \,\frac{\omega_\ast^{D-4}}{D-4}\oint F^\ast_{\mu\nu}(1,\hat x) P^{\mu\nu,\rho\sigma}F_{\rho\sigma}(1,\hat x)\,\frac{d\Omega(\hat x)}{2(2\pi)^{D-1}}
\end{equation}
and, to leading order in $\epsilon = (4-D)/2$, 
\begin{equation}\label{}
	\label{eq:Imdeltaomegaanglesepsilon}
	\operatorname{Im}2\delta(\sigma,b) = -\frac{\omega_\ast^{-2\epsilon}}{4\hbar\epsilon}\oint F^\ast_{\mu\nu}(1,\hat x) P^{\mu\nu,\rho\sigma}F_{\rho\sigma}(1,\hat x)\,\frac{d\Omega(\hat x)}{2(2\pi)^3}\,.
\end{equation}

\subsection{Infrared divergences in the number of emitted quanta}

The eikonal operator \eqref{eq:eiksr} is manifestly unitary (once we neglect the quantum remainder $\Delta$). As such, it better yield 1 when we insert $\mathcal O=1$ in Eq.~\eqref{SOS}.
Indeed, we can apply Eq.~\eqref{eq:ON} with $\mathcal O=1$ to compute the probability $\mathcal P_N$ for the emission of $N$ gravitons:
\begin{equation}\label{}
	\mathcal P_N  = \langle 1\rangle_N\,,
	\qquad 
	\langle 1 \rangle= 
	\sum_{N=0}^\infty \mathcal P_N=1\,.
\end{equation} 
We thus need to evaluate
\begin{equation}\label{calPN}
	\mathcal P_N =
	\frac{1}{N!}\,
	\sum_{j_1,\ldots,j_N} \int_{\vec k_1}\cdots \int_{\vec k_N}\,
	\langle 0|S^\dagger_{s.r.} a_{j_1}^\dagger\cdots a_{j_N}^\dagger |0\rangle
	\langle 0|a_{j_1}\cdots a_{j_N} S_{s.r.} |0\rangle\,.
\end{equation}
Each $a_j(k)$ (resp. $a^\dagger_j(k)$) oscillator pulls down a factor $f_j$ (resp. $f_j^\ast$) so that
\begin{equation}\label{}
	\mathcal P_N 
	=
	\frac{1}{N!}
	\langle 0 |S^\dagger_{s.r.}|0\rangle 
	\left[
	\frac1{\hbar}
	\int_{\vec k}F^\ast_{\mu\nu}(k)P^{\mu\nu,\rho\sigma}F_{\rho\sigma}(k)
	\right]^N 
	\langle 0 |S_{s.r.}|0\rangle \,.
\end{equation}
Recognizing the same integral as the one appearing in \eqref{eq:Imdeltadef}, up to a crucial factor of $2$, and using the fact that $\langle 0| S_{s.r.}|0\rangle= e^{i2\delta}$ as in \eqref{expi2delta}, we thus have
\begin{equation}\label{}
	\mathcal P_N 
	=
	\frac{1}{N!} 
	\left[
	2\operatorname{Im}2\delta
	\right]^N 
	e^{-2\operatorname{Im}2\delta}\,.
\end{equation}
In this way we obtain that the probability for the emission of $N$ gravitons follows a Poisson distribution
with $2\operatorname{Im}2\delta(\sigma,b)$ the average number of emitted gravitons  
\begin{equation}
	\sum_{N=0}^{\infty} N \mathcal P_N= 2\operatorname{Im}2\delta \,.
	\label{BN3}
\end{equation}
Again let us note that
$2\operatorname{Im}2\delta$ is divergent both as $D\to4$ and as $\hbar \rightarrow 0$.

Alternatively, one can directly insert in \eqref{SOS} the operator
\begin{equation}\label{Number}
	N = \int_{\vec k}\sum_j a^\dagger_{j}(k) a_{j}(k)\,,\qquad 
	\mathcal N = \langle N \rangle
	\,,
\end{equation}
which counts the number of emitted gravitons.
Consistently with \eqref{BN3}, in this way one finds
\begin{equation}\label{}
	\mathcal N  = \frac1{\hbar} \int_{\vec k}F^\ast_{\mu\nu}(k)P^{\mu\nu,\rho\sigma}F_{\rho\sigma}(k)
	= 2\operatorname{Im}2\delta\,.
\end{equation}

Of course, a similar discussion applies for the number of emitted quanta in supergravity. There, one simply obtains a product of Poisson distributions each associated with a species of emitted particle. Similarly, one can consider insertion of number operators vectors and scalars as well.
In each case, such expectation values coincide with twice \eqref{Imdeltavec} and \eqref{Imdeltasc}.

\subsection{Zero-frequency limit (ZFL) of  \texorpdfstring{$\frac{dE_{\text{rad}}}{d\omega}$}{dE/dw}}

We now move on to the insertion of the energy-momentum operator
\begin{equation}\label{}
	P^\alpha 
	= 
	\int_{\vec k}\sum_j \,\hbar k^\alpha\, a^\dagger_{j}(k) a_{j}(k)\,,
	\qquad
	\mathcal P^\alpha_{\text{rad}} = \langle P^\alpha \rangle
\end{equation}
as in Eq.~\eqref{SOS}, which leads to 
\begin{equation}
	\label{eq:Erad}
	\mathcal P^\alpha_{\text{rad}} = \int_{\vec k}  k^\alpha F_{\mu\nu}(k) P^{\mu\nu,\rho\sigma} F_{\rho\sigma}(k) \,,
\end{equation}
thanks to \eqref{cancellation}.
Focusing on $D=4$ from now on for this quantity, and making the dependence on the upper cutoff $\omega_\ast$ explicit, we thus consider
\begin{equation}\label{}
	\mathcal P^\alpha_\text{rad}(\omega_\ast) = \int_{\vec k}\theta(\omega_\ast-k^0)\,k^\alpha F_{\mu\nu}(k) P^{\mu\nu,\rho\sigma} F_{\rho\sigma}(k)
\end{equation}
and study the soft emission spectrum i.e. the derivative
\begin{equation}\label{Prad}
	\frac{d\mathcal P_\text{rad}^\alpha(\omega_\ast)}{d\omega_\ast} =\int_{\vec k}\delta(\omega_\ast-k^0)\,k^\alpha F_{\mu\nu}(k) P^{\mu\nu,\rho\sigma} F_{\rho\sigma}(k)
\end{equation}
as $\omega_\ast\to0$. For later convenience we also rewrite \eqref{Prad} introducing explicitly the integration over the frequency and the angles, using $k^\mu = \omega(1,\hat x)$ and $F^{\mu\nu}(k) = \omega^{-1}F^{\mu\nu}(1,\hat x)$,
\begin{equation}\label{dPdomega}
	\frac{d\mathcal P^\alpha_\text{rad}(\omega_\ast)}{d\omega_\ast} =\oint (1,\hat x)^\alpha F_{\mu\nu}(1,\hat x) P^{\mu\nu,\rho\sigma} F_{\rho\sigma}(1,\hat x) \, \frac{d \Omega(\hat{x})}{2(2 \pi)^{3}}\,.
\end{equation}
Note that under a Lorentz transformation $p^\mu \to p'^\mu=(\Lambda^{-1})\indices{^\mu_\nu}p^\nu$, 
\begin{equation}\label{Lorentzbehavior}
	\begin{aligned}
		\frac{dP_\text{rad}^\alpha(\omega_\ast)}{d\omega_\ast} &\to \int_{\vec k}\delta(\omega_\ast-(\Lambda k)^0)\,k^\alpha F_{\mu\nu}(k) P^{\mu\nu,\rho\sigma} F_{\rho\sigma}(k)\\
		&=
		\oint \frac{\Lambda\indices{^\alpha_0}+\Lambda\indices{^\alpha_I} \hat{x}^{I}}{\Lambda\indices{^0_0}+\Lambda\indices{^0_J}\hat{x}^{J}}\, F_{\mu\nu}(1,\hat x) P^{\mu\nu,\rho\sigma} F_{\rho\sigma}(1,\hat x)\, \frac{d \Omega(\hat{x})}{2(2 \pi)^{3}}\,.
	\end{aligned}
\end{equation} 
Therefore, while $\frac{d\mathcal P^0_{\text{rad}}}{d\omega_\ast}=\frac{dE_{\text{rad}}}{d\omega_\ast}$ is Lorentz invariant, the spatial  components $\frac{d\mathcal P^I_{\text{rad}}}{d\omega_\ast}$ have more complicated transformation laws.

We thus focus on the ZFL of the energy emission spectrum $\frac{dE_{\text{rad}}}{d\omega_\ast}$, i.e.~the $\alpha=0$ component of \eqref{dPdomega}, which reads
\begin{equation}\label{dEdomega}
	\frac{dE_\text{rad}(\omega_\ast)}{d\omega_\ast} =\oint F_{\mu\nu}(1,\hat x) P^{\mu\nu,\rho\sigma} F_{\rho\sigma}(1,\hat x) \, \frac{d \Omega(\hat{x})}{2(2 \pi)^{3}}\,,
\end{equation}
 while deferring to Appendix~\ref{app:dPI} an analysis of the spatial components. 
Comparing \eqref{dEdomega} with \eqref{eq:Imdeltaomegaanglesepsilon}, we immediately see that they are identical up to the prefactor $-4\hbar\epsilon$, so that 
\begin{equation}
	\label{eq:ZFLspectrum}
	\lim_{\omega \to0}\frac{dE_{\text{rad}}}{d\omega} = \lim_{\epsilon\to 0} \left[-4\hbar\epsilon \operatorname{Im}2\delta(\sigma,b) \right].
\end{equation}
This highlights a general mechanism: the infrared divergences of the elastic amplitude determine the ZFL of the energy emission spectrum via massless quanta \cite{Weinberg:1965nx,DiVecchia:2021ndb}.
Eq.~\eqref{eq:ZFLspectrum} generalizes trivially to an arbitrary background process $\alpha\to\beta$ and captures the exact dependence on the associated kinematics, regardless whether the involved particles carry spin.
Similar links hold between the energy spectra for massless vector and scalar emissions and \eqref{Imdeltavec}, \eqref{Imdeltasc}, and between the total emission spectrum in $\mathcal N=8$ and \eqref{N8m}.

\section{ZFL (and \texorpdfstring{$\operatorname{Im}2\delta$}{Im2d}) at Arbitrary Velocities}
\label{sec:aboveandbelow}

In this section we provide explicit expressions for $\operatorname{Im}2\delta$ and $\frac{dE_{\text{rad}}}{d\omega}$, specialize them to the case of gravitational $2\to2$ scattering and analyse their properties as the relative velocity of the colliding objects varies. As we already discussed for the memory waveform, we will observe a non-trivial transition between the standard PM regime where $Q\ll m_1, m_2$ and the region characterized by the bound \eqref{eq:KTb}.

The integrals entering \eqref{eq:Imdeltadef} can be evaluated to leading order in $\epsilon = (4-D)/2$ using the basic identity (see \cite{Weinberg:1965nx} and Appendix~B of \cite{Heissenberg:2021tzo})
\begin{equation}\label{eq:basi}
	\int \frac{d^{4-2\epsilon}k}{(2\pi)^4} \frac{2\pi\delta(k^2)\theta(k^0)\theta(\omega_\ast-k^0)}{(p_n k)(p_m k)}
	= -
	\frac{\omega_\ast^{-2\epsilon}}{8\pi^2\epsilon}\,\frac{F_{nm}}{m_n m_m}\,,
\end{equation}
where
\begin{equation}\label{keyf}
	F_{nm}=\frac{\eta_n \eta_m \operatorname{arccosh}{\sigma_{nm}}}{ \sqrt{\sigma_{nm}^2-1}}
	\,,\qquad
	\sigma_{nm}=-\eta_n\eta_m \, \frac{p_n\cdot p_m}{m_n m_m}\,.
\end{equation}
Note that, despite the presence of a cutoff, the left-hand side of \eqref{eq:basi} is Lorentz-invariant to leading order in $\epsilon$, and indeed the $1/\epsilon$ pole on the right-hand side is $\omega_\ast$-independent (see also the discussion around \eqref{Lorentzbehavior}).
Then, to leading order in the limit $\epsilon\to 0$, Eq.~\eqref{eq:Imdeltadef} together with \eqref{eq:ZFLspectrum} gives:
\begin{equation}
	\lim_{\omega \to0}\frac{dE^{\text{gr}}}{d\omega} = \frac{2G }{\pi} \sum_{n,m} m_n m_m  \left(\sigma_{nm}^2 - \frac{1}{2} \right) F_{nm}\,.
	\label{Imdelta}
\end{equation}
Let us emphasize that in this expression the dependence on the kinematics of elastic process is exact. Like the soft theorem, this formula is insensitive to the specific details of the hard particles and should also hold if they carry spin. Moreover, as expected, $\operatorname{Im}2\delta>0$ for $\epsilon<0$, which grants the convergence of the integral, so that $e^{-\operatorname{Im}2\delta}$ is indeed an exponential suppression: this factor tends to zero as $D\to4$ and as $\hbar \to 0 $, indicating that $|0\rangle$ and $S_{s.r.}|0\rangle$ have zero overlap in these limits. Correspondingly, via \eqref{eq:ZFLspectrum}, the ZFL of $dE/d\omega$ is also positive.

The right-hand side of Eq.~\eqref{Imdelta} generalizes straightforwardly to generic background processes $\alpha\to\beta$ involving an arbitrary number of (massive and massless) states. Again via \eqref{eq:ZFLspectrum}, it is related to the quantity to be exponentiated in momentum space in order to resum all infrared divergences in the exclusive amplitude $\mathcal A_{\alpha\beta}$ due to soft virtual gravitons \cite{Weinberg:1965nx,Heissenberg:2021tzo}: $\mathcal A_{\alpha\beta} = e^{\mathcal W_{\alpha\beta}} \mathcal A_{\alpha\beta}^0$, where $\mathcal A_{\alpha\beta}^0$ is infrared-finite and
\begin{equation}\label{}
	\mathcal W_{\alpha\beta} = \frac{\kappa^2 \omega_\ast^{-2\epsilon} }{ (4\pi)^2 \hbar\epsilon} \sum_{n,m} m_n m_m  \left(\sigma_{nm}^2 - \frac{1}{2} \right) \frac{\eta_n \eta_m\operatorname{arccosh}{\sigma_{nm}}-i\pi \eta_{nm}}{\sqrt{\sigma^2_{nm} -1}}
\end{equation}
with $\eta_{nm}=\delta_{\eta_n,\eta_m}-\delta_{nm}$. In particular,
\begin{equation}\label{WImdeltalink}
	\operatorname{Im}2\delta = -\operatorname{Re}\mathcal W 
\end{equation}
for the $2\to2$ process which is the object of our main interest here.

The discussion presented so far can be straightforwardly generalised to ${\cal N}=8$ supergravity starting from the soft eikonal operator~\eqref{eq:dn8sr}. The technical steps are identical to the GR case, except that one needs to add all  contributions to $\operatorname{Im}2\delta$ that arise by reordering the ladder operators associated to the various massless particles in the theory, which include vectors and scalars. Thus, in analogy with \eqref{Imdelta}, we find: 
\begin{align}\label{}
	\lim_{\omega \to0}\frac{dE^{\text{vec}}}{d\omega} &= \frac{1}{4 \pi^2} \sum_{n,m} e_n e_m (-\sigma_{nm}) F_{nm}\,,
	\label{Imdeltavec}
	\\
	\lim_{\omega \to0}\frac{dE^{\text{sc}}}{d\omega} &= \frac{1}{4 \pi^2}   \sum_{n,m} \frac{g_n g_m}{m_nm_m} F_{nm}\,.
	\label{Imdeltasc}
\end{align} 
After summing up the all such terms related to the various physical modes we obtain a remarkably simple result:
\begin{equation}\label{N8m}
\lim_{\omega \to 0}\frac{dE^{\mathcal N=8}}{d\omega}
	=
	 \frac{2G }{\pi}
	\sum_{n,m}
	m_n m_m (\sigma'_{nm})^2 F_{nm}\,,
\end{equation} 
where $\sigma_{nm}'=\sigma_{nm}-1$ if $n$ and $m$ have momenta compactified along the same KK direction (so that $m_n=m_m$) and  $\sigma_{nm}'=\sigma_{nm}$ otherwise. Again, the right-hand side of \eqref{N8m} is related to the exponentiation of infrared divergences due to soft graviton, dilatons and massless KK modes,
\begin{equation}\label{WN8}
	\mathcal W^{\mathcal N=8}_{\alpha\beta}
	=
	\frac{\kappa^2 \omega_\ast^{-2\epsilon}}{(4\pi)^2 \hbar \epsilon}
	\sum_{n,m}
	m_n m_m (\sigma'_{nm})^2 \frac{\eta_n \eta_m\operatorname{arccosh}{\sigma_{nm}}-i\pi\eta_{nm}}{\sqrt{\sigma^2_{nm} -1}}
\end{equation}
by the same link as in the case of graviton emissions \eqref{WImdeltalink}. 

Let us now study more in detail the spectrum for the $2\to2$ process in General Relativity.
Starting from the general expression \eqref{Imdelta} for $\operatorname{Im}2\delta$, it is sufficient to use $\sigma_{nn}=1$ and $F_{nn} = 1$, while for $n\not= m$ we have $\sigma_{nm}=\sigma_{mn}$ and
\begin{equation}\label{eq:sigmamn4p}
	\sigma_{12} = \sigma_{34} = \sigma\,,
	\quad
	\sigma_{13} = \sigma_{24} = \sigma_Q\,,\quad
	\sigma_{14} = 1+\frac{Q^2}{2m_1^2}\,,\quad
	\sigma_{23} = 1+\frac{Q^2}{2m_2^2}\,,
\end{equation}
where we introduced the shorthand notation
\begin{equation}\label{eq:sigmaq}
	\sigma_Q = \sigma -\frac{Q^2}{2m_1m_2} = - \frac{u - m_1^2 - m_2^2}{2 m_1m_2}\,.
\end{equation}
Then, Eq.~\eqref{Imdelta} becomes 
\begin{align}
\nonumber
		\lim_{\omega \to0}\frac{dE^{\text{gr}}}{d\omega} &=  \frac{4 G }{\pi}
		\Bigg[
		2m_1m_2\left(\sigma^2-\tfrac12\right)\frac{\operatorname{arccosh}\sigma}{\sqrt{\sigma^2-1}}
		-
		2m_1m_2\left(\sigma_Q^2-\tfrac12\right)\frac{\operatorname{arccosh}\sigma_Q}{\sqrt{\sigma_Q^2-1}}
		\\  \label{eq:d2gr}
		&+\frac{m_1^2}{2}
		-
		m_1^2 \Big(\left(
		1+\tfrac{Q^2}{2m_1^2}
		\right)^2-\tfrac12\Big)
		\frac{\operatorname{arccosh}\left(
			1+\tfrac{Q^2}{2m_1^2}
			\right)}{\sqrt{\left(
				1+\tfrac{Q^2}{2m_1^2}
				\right)^2-1}}
		\\ \nonumber
		&
		+\frac{m_2^2}{2}
	-
		m_2^2 \Big(\left(
		1+\frac{Q^2}{2m_2^2}
		\right)^2-\tfrac12\Big)
		\frac{\operatorname{arccosh}\left(
			1+\tfrac{Q^2}{2m_2^2}
			\right)}{\sqrt{\left(
				1+\tfrac{Q^2}{2m_2^2}
				\right)^2-1}}
		\Bigg]\,,
\end{align} 
where, as discussed, the transferred momentum $Q$ should be interpreted by using~\eqref{qtoQ2}. As already emphasized, while in the following we will focus on certain interesting kinematic limits, the dependence of this formula on the dynamics of the background elastic process, and in particular on $Q/m_i$, is exact.

The standard PM regime considered in~\cite{Damour:2020tta,DiVecchia:2021ndb} requires that 
\begin{equation}\label{}
	Q^2 \sim (p \Theta_s)^2 \ll 2m_i^2\,.
\end{equation}
In this regime, one can extract the leading (3PM) contribution by Taylor-expanding the first line of~\eqref{eq:d2gr} in $Q^2$, while the remaining two lines only give subleading contributions,
\begin{equation}
  \label{eq:standpm}
 \lim_{\omega \to0}\frac{dE^{\text{gr}}}{d\omega}  \simeq  \frac{2G }{\pi} Q^2
	\left[
	\frac{8-5\sigma^2}{3(\sigma^2-1)}
	+
	\frac{(2\sigma^2-3)\sigma\operatorname{arccosh}\sigma}{(\sigma^2-1)^{3/2}}
	\right].
\end{equation}
Note that, in the UR regime,  this gives:
\begin{equation}
  \label{eq:standpmur}
 \lim_{\omega \to0}\frac{dE^{\text{gr}}}{d\omega}  \simeq  \frac{4G }{\pi} Q^2  \left( \log\frac{s}{m_1 m_2} - \frac56 \right).
\end{equation}
When rewritten in terms of $\operatorname{Im}2\delta(\sigma,b)$, the result \eqref{eq:standpm} agrees with Eq.~(5.14) of~\cite{DiVecchia:2021ndb} once we use $Q\simeq p\, \Theta_s$ and the leading result~\eqref{eq:Theta1PM} for $\Theta_s$.
Moreover, to the leading 3PM order, Eq.~\eqref{eq:standpm} has been explicitly shown to hold also if the colliding objects  carry spin, for generic spin alignments \cite{Alessio:2022kwv}.  Via unitarity, analyticity and crossing symmetry, the 3PM divergent part of $\operatorname{Im}2\delta$ immediately provides the radiation-reaction corrections to the 3PM deflection \cite{DiVecchia:2021ndb,DiVecchia:2021bdo}.

Still, let us stress that the exact dependence of~\eqref{eq:d2gr} on $Q^2$ can be used to extract a prediction for the ZFL of the spectrum and, via \eqref{eq:ZFLspectrum}, for the IR-divergent part of $\operatorname{Im}2\delta$ also at higher orders in $G$. In particular, once the $n$PM deflection angle is known, Eq.~\eqref{eq:d2gr} provides the IR divergent part of $2\delta$ at $(n+2)$PM order (so far explicit results for $\Theta_s$, including radiation-reaction effects, are available up to 3PM).

Notice that~\eqref{eq:standpm} is determined by the soft theorems and the $1$PM deflection angle, explaining why the probe limit $\nu=(m_1 m_2)/(m_1+m_2)^2\ll 1$ used in~\cite{Smarr:1977fy} captures the full result at this order. It is straightforward to use~\eqref{eq:d2gr} and extract the higher PM corrections to~\eqref{eq:standpm} and so to the ZFL of the energy spectrum. Of course starting at $3$PM order one finds a non-trivial dependence on $\nu$ inherited from the expression of $\Theta_s$.

As a check, we compared the small-velocity limit of the first three orders in the PM expansion of~\eqref{eq:d2gr} with the results obtained in the standard Post-Newtonian (PN) approach, see for instance the results in Appendix A of~\cite{Bini:2021jmj}.\footnote{Starting from (A6) and (A7) in that reference, it is straightforward to perform the ZFL of the terms that do not receive contributions from the integrals in (A8) and see that they agree with the PN expansion of our~\eqref{eq:d2gr}. One can perform a more detailed comparison verifying that all terms agree at the 2PN level considered in~\cite{Bini:2021jmj}. We thank Donato Bini for performing this check.} It is interesting to notice that the radiation-reaction contribution to the deflection angle~\cite{DiVecchia:2020ymx} yields terms in the soft spectrum that have odd powers of the velocity in the PN expansion. The first such contribution to the ZFL of the energy spectrum can be obtained simply by inserting in~\eqref{eq:standpm} the 2.5PN term of the deflection angle in GR~\cite{Damour:2020tta,DiVecchia:2021ndb,DiVecchia:2021bdo} and is $\frac{dE^{\text{gr}}}{d\omega}\sim\frac{512 G^5 \nu^3 m^6 v}{25 \pi  b^4}$. At this order one would expect also a contribution from the cross term of the linear and non-linear memories of the waveforms, however, by using the result of~\cite{Wiseman:1991ss}, one can check that such contribution vanishes after integration over the angles at least at the leading PN order. Thus the result quoted above should be the full 5PN correction to the leading ZFL of the energy spectrum and it would be interesting to compare it with the results derived in the PN approach.

In the naive small-$\Theta_s$ expansion, including the leading term in~\cite{Smarr:1977fy}, there are terms that are logarithmically divergent as $\sigma \to \infty$ with $\Theta_s$ fixed, i.e.~when one enters the kinematic regime defined in~\eqref{eq:KTbex}. 
However, as discussed in the previous section, the PM approximation can break down even when $\Theta_s$ is small, see~\eqref{eq:KTb}.
This happens when\footnote{Note that this condition requires at least one of the two particles to be relativistic, $p \gg m_i$.}, for at least one index $i$,
\begin{equation}
  \label{eq:KTbex}
  \frac{Q}{\sqrt{2}\, m_i} =  \frac{\sqrt{2}\,p}{m_i} \sin\frac{\Theta_s}{2} \sim \frac{G (s- m_1^2-m_2^2)}{m_i b} \gtrsim 1 
\end{equation}
In this regime, one cannot expand the last two lines in~\eqref{eq:d2gr} for small $\Theta_s \sim G$.

The first line in~\eqref{eq:d2gr}  can always  be expanded to first order in $Q^2/s$ (and exhibits a singular massless limit) while the second and third line depend very non-trivially on the ``scaling variables" $\xi_1 \equiv \frac{Q}{m_1},\xi_2 \equiv \frac{Q}{m_2}$, respectively. 
Actually, when seen as analytic functions of $z = \xi^2$, those two lines exhibit a branch point on the negative real axis at the unphysical point $z = -4$, corresponding to the $t$-channel thresholds $t = 4 m_i^2$. This implies that the PM expansion around $z=0$ starts to diverge at $\xi_i^2 = 4$. This quantifies the qualitative statement made in \eqref{eq:KTbex}.

This complicated $\xi_i$-dependence smoothly connects several interesting regimes. We have already discussed the conventional 3PM regime where \eqref{eq:standpm} holds. 
At the opposite end let's consider  the extreme ultrarelativistic regime, or equivalently the massless limit, where $2p \to \sqrt{s}$ and $m_1, m_2\ll  Q = \sqrt{s} \sin\frac{\Theta_s}{2}$. The mass singularities neatly cancel and then~\eqref{eq:d2gr} reduces to (see e.g.~\cite{Addazi:2019mjh} where the result is extended to an arbitrary number of external massless legs) 
\begin{equation}\label{eq:urd2}
	 \lim_{\omega \to0}\frac{dE^{\text{gr}}}{d\omega} \simeq  \frac{4 G }{\pi}	\left[
	s \log \frac{s}{s-Q^2}
	+
	Q^2\log\frac{s-Q^2}{Q^2}
	\right]_{Q=\sqrt{s} \sin\frac{\Theta_s}{2}}\,,
\end{equation}
which also affords a very compact form in terms of the deflection angle
\begin{equation}\label{yr}
 \lim_{\omega \to0}\frac{dE^{\text{gr}}}{d\omega} \simeq - \frac{4 G }{\pi}s \left[\cos^2 \frac{\Theta_s}{2} \log \cos^2 \frac{\Theta_s}{2} + \sin^2 \frac{\Theta_s}{2} \log \sin^2 \frac{\Theta_s}{2} \right].
\end{equation}
which agrees with the leading soft limit of Eq.~(5.12) of$\,$\footnote{In order to reinstate the Newton constant, that result should be multiplied by $8\pi G$ and in that equation $E=\sqrt{s}/2$ indicates the energy of each incident particle.}\;\cite{Sahoo:2021ctw}. Let us consider the small $\Theta_s$ limit of~\eqref{yr}. At leading order for $\Theta_s\ll 1$ we have
\begin{equation}
  \label{eq:lm0}
  \lim_{\omega_\ast\to0}
  	\frac{dE_{\text{rad}}}{d\omega_\ast} \simeq \frac{G s \Theta_s^2}{\pi} \left[1 + \log\frac{4}{\Theta_s^2}
	\right],
\end{equation}
which reproduces the result obtained in~\cite{Gruzinov:2014moa} within a classical GR approach and in~\cite{Ciafaloni:2018uwe} from a scattering amplitudes perspective. 

The latter approach clarifies the origin of the non-analytic behaviour in $G$ in~\eqref{eq:lm0} as we summarise below. Most of the radiated energy due to soft modes is in a region almost collinear to the hard particles as it is clear from~\eqref{eq:wavAetm0} and~\eqref{eq:wavAepm0} since the denominators in those expressions become smaller as $\theta\ll 1$ or $\pi-\theta\ll 1$. However, when $\theta \lesssim \Theta_s$ ($\pi-\theta \lesssim \Theta_s$) the waveforms stop growing and there is a plateau till $\theta=0$ ($\theta=\pi$). The key point is that, during the scattering, the direction of motion of the hard particles changes classically, so the radiation collinear with the initial states has an angle $\Theta_s$ with the final states and vice versa. This misalignment regulates the integral over $\theta$ in the energy spectrum and this produces the non-analytic dependence on $G$ in~\eqref{eq:lm0}.

 It is then clear that it is essential to keep track of the (elastic) eikonal exponentiation to encode the information about the classical deflection angle, see the discussion starting from~\eqref{eq:deltaQ0}. Technically, this is done by using the soft eikonal operator~\eqref{eq:eiksr}, where the operator part describing the radiation acts on the full elastic eikonal, instead of just on the amplitude describing a single graviton exchange (which is a quantum process). In the standard relativistic PM regime ({\em i.e.} away from the regime \eqref{eq:KTbex}) the subtlety discussed here is not manifest, since in that case the would-be collinear blow up in~\eqref{eq:wavAetPM1} and~\eqref{eq:wavAepPM1} is cut-off by $E_i/p -1 > 0$. As a result  the $\theta$ integral in the spectrum is regulated by $\sigma$ as it is clear from~\eqref{eq:standpm}. Eq.~\eqref{eq:d2gr} provides a smooth transition between the two regimes and the $\log\sigma$ enhancement in~\eqref{eq:standpm} becomes $\log\Theta_s^{-2}$ when entering the extreme ultra-relativistic region.

Let us finally consider another corner of our two-parameter space, the one in which, say, $\xi_1=\frac{Q}{m_1}\to 0$ and $\xi_2 =\frac{Q}{m_2}\to \infty$. To this end, we can first take $m_2$ to be small for fixed $s$, $m_1$, $Q$, and then take the small-$Q$ limit for fixed $s$, $m_1$. This time the singularity for $m_2 \to 0$ cancels while a $\log m_1$ remains, so that:
\begin{equation}
\label{m1ggm2}
 \lim_{\omega \to0}\frac{dE^{\text{gr}}}{d\omega} \simeq \frac{G}{3 \pi} Q^2 \left[1+ 6 \log\left(  \frac{(s-m_1^2)^2}{m_1^2 Q^2} \right)\right] .
\end{equation}
Inside this regime we can also consider the probe limit, where we can regard particle 1 as the primary object and particle 2 as a test mass. This case corresponds to
further taking, in \eqref{m1ggm2}, $s\simeq m_1^2+2m_1 p$ (i.e.~$p\simeq m_2\sigma$) obtaining
\begin{equation}
	\label{m1ggm2exp}
	\lim_{\omega \to0}\frac{dE^{\text{gr}}}{d\omega} \simeq \frac{G}{3 \pi} Q^2 \left[1+ 6 \log\left(  \frac{4p^2}{Q^2} \right)\right] ,
\end{equation}
or, recalling $Q = 2p\sin\frac{\Theta_s}{2}$,
\begin{equation}
  \label{eq:smac}
  \lim_{\omega \to 0}\frac{dE_{\text{rad}}}{d\omega} \simeq \frac{4 G p^2 \sin^2\frac{\Theta_s}{2}  \left[1-6 \log \left(\sin^2\frac{\Theta_s}{2} \right)\right] }{3 \pi }\,,
\end{equation}
where here $\Theta_s$ is the deflection angle in the probe limit in a Schwarzschild black hole of mass $m_1$. 

Let us now turn to the case of ${\cal N}=8$ supergravity.
Specializing \eqref{N8m} to the $2\to2$ kinematics as in \eqref{eq:sigmamn4p}, we have
\begin{align}
	\label{eq:d2n8}
	\lim_{\omega \to0}\frac{dE^{{\cal N}=8}}{d\omega} &=  \frac{4 G }{\pi} 
	\Bigg[
	2m_1m_2 \sigma^2\, \frac{\operatorname{arccosh}\sigma}{\sqrt{\sigma^2-1}}
	-
	2m_1m_2 \sigma_Q^2\, \frac{\operatorname{arccosh}\sigma_Q}{\sqrt{\sigma_Q^2-1}}
	\\ \nonumber
	&-
	\frac{(Q^2)^2}{4 m_1^2}\,
	\frac{\operatorname{arccosh}\left(
		1+\tfrac{Q^2}{2m_1^2}
		\right)}{\sqrt{\left(
			1+\tfrac{Q^2}{2m_1^2}
			\right)^2-1}}
	-
	\frac{(Q^2)^2}{4 m_2^2}\,
	\frac{\operatorname{arccosh}\left(
		1+\tfrac{Q^2}{2m_2^2}
		\right)}{\sqrt{\left(
			1+\tfrac{Q^2}{2m_2^2}
			\right)^2-1}}
	\Bigg]_{Q=2 p \sin\frac{\Theta_s}{2}}\,.
\end{align}
The standard relativistic regime, where $Q^2\ll 2m_i^2$ in the equation above, has an analytic PM expansion whose leading term reproduces the result of~\cite{DiVecchia:2021bdo}
\begin{equation}
	\label{eq:standpmn8}
	\lim_{\omega \to0}\frac{dE^{{\cal N}=8}}{d\omega} \simeq \frac{4G\, Q^2}{\pi} 
	\left[
	\frac{\sigma^2}{\sigma^2-1}
	+
	\frac{(\sigma^2-2)\sigma\operatorname{arccosh}\sigma}{(\sigma^2-1)^{3/2}}
	\right],
\end{equation}
where the leading deflection angle is given in~\eqref{eq:Theta1PM} now with $\zeta=0$. On the contrary, in the regime~\eqref{eq:KTbex}, the small-$\Theta_s$ expansion is non-analytic and interestingly, in extreme ultrarelativistic kinematics where the masses can be neglected, one obtains again~\eqref{eq:urd2}. Indeed, the contributions related to the dilaton, and the Kaluza-Klein scalars and vectors become negligible in this regime, as suggested by the fact that \eqref{Imdeltavec}, \eqref{Imdeltasc} scale with lower powers of $\sigma_{nm}$ compared to \eqref{Imdelta}, and the graviton provides the dominant, universal behaviour.

We refrain here from discussing in detail the soft spectra of $\mathcal N=8$ supergravity. Suffice it to say that the link \eqref{eq:ZFLspectrum} trivially holds for each ``species'' of radiated particle (dilaton, KK modes and gravitons), so that each spectrum is determined by the corresponding divergent part of $\operatorname{Im}2\delta$ discussed in the previous section.
In particular, the ZFL of the full ${\cal N}=8$ energy spectrum, obtained by summing over all types of emission, is obtained by substituting \eqref{eq:d2n8} (or more generally \eqref{N8m}) in~\eqref{eq:ZFLspectrum}. At ultra-high energies, only the contribution due to the emission of gravitons survives. This is a universal expression for two-derivatives theories in accordance with the expectation that gravity dominates the high-energy limit not just for the elastic scattering, as argued in~\cite{DiVecchia:2020ymx}, but also in the radiation sector. 

\section{Discussion and Outlook}
\label{sec:disc}

In this paper we focused on the soft eikonal operator describing the emission of low-frequency gravitons (or, in general, massless states in ${\cal N}=8$ supergravity) from the $2\to 2$ scattering of energetic particles. By combining Weinberg's exponentiation in momentum space and the eikonal exponentiation in impact parameter space, we obtained explicit formulae for the waveforms and the energy spectrum in the zero frequency limit. The main feature is that these observables are smooth as the energy of the collision increases and display a qualitative change in their behaviour when one goes above the threshold~\eqref{eq:KTb}. In general, waveforms and spectra now depend non-trivially on the two ratios appearing in~\eqref{eq:KTb} if they are kept fixed as one takes $s \to \infty$. In  the extreme ultrarelativistic regime in which {\it both} ratios go to infinity (e.g.~in the massless case), the universality of gravitational scattering~\cite{DiVecchia:2020ymx} is restored also for radiative observables at least in the ZFL regime: for instance, the dilaton waveform becomes negligible (see Eq.~\eqref{eq:Wdil}, while in the standard PM region it is non-trivial~\eqref{eq:Wdils}) and the energy flux reproduces the universal result~\eqref{eq:lm0}. Note that, in the standard PM regime,~\eqref{eq:standpmur} shows that the ZFL flux divided by the initial energy increases logarithmically as $\sigma$ becomes larger: $\frac{1}{\sqrt{s}} \frac{dE_\text{rad}}{d\omega_\ast} \sim G\sqrt{s}\, \Theta_s^2 \log\sigma$. Even such a mild increase is inconsistent at high energy and indeed when the threshold~\eqref{eq:KTb} is crossed the behaviour of the spectrum smoothly changes to~\eqref{eq:lm0}: the logarithmic increase with the energy is substituted by a non-analytic dependence on the scattering angle (and thus on the Newton constant).

Of course, the immediate next challenge is to extend our approach beyond the small frequency limit. It is possible to introduce an eikonal operator that extends~\eqref{eq:Asr} to generic values of the frequency and assume an exponentiation in impact parameter space as done for the elastic eikonal. A first proposal in this direction has been discussed in detail in~\cite{Cristofoli:2021jas} which focuses on the regime below the threshold~\eqref{eq:KTb}. In this case the operator part of the eikonal operator is directly related to the leading PM waveforms~\cite{Kovacs:1978eu,Jakobsen:2021smu,Mougiakakos:2021ckm,Riva:2021vnj}. Various checks on the consistency of this proposal are already discussed in the same reference and it is possible~\cite{tap1} to use it as a starting point to discuss the qualitative features of the energy spectrum (for both $dE_{\rm rad}/d\omega_\ast$ and the fully differential $dE_{\rm rad}/(d\omega_\ast \,d\Omega)$) along the lines of what was done in the massless case~\cite{Gruzinov:2014moa,Ciafaloni:2015xsr,Ciafaloni:2018uwe}. The low frequency approximation we considered here certainly breaks down when $\omega b\simeq 1$ and so one can estimate the contribution of the very soft gravitons to the total energy radiated simply by assuming that the spectrum is constant up to $\omega \sim 1/b$. Then the contribution from soft gravitons is $E_{\rm rad} \sim \sqrt s\,\Theta_s^3$, which is of the same order as the full result~\cite{Herrmann:2021lqe,Herrmann:2021tct} in the PN region $\sigma \sim 1$. When $\sigma \gg 1$, but still below the threshold~\eqref{eq:KTb}, $E_{\rm rad}$ scales as $E_{\rm rad} \sim  \sqrt s\,\Theta_s^3 \sqrt{\sigma}$~\cite{Herrmann:2021lqe,Herrmann:2021tct}, while the prediction from~\cite{Gruzinov:2014moa,Ciafaloni:2015xsr,Ciafaloni:2018uwe} for the massless case (which of course is above the threshold~\eqref{eq:KTb}) is $E_{\rm rad} \sim  \sqrt s\, \Theta_s^2 \log\Theta_s^{-2}$. It will be interesting to check whether these two cases (massive and massless) are smoothly connected by taking the extreme ultrarelativistic limit, as it happens for the ZFL case. The power-like dependence on the scattering angle is consistent with the pattern seen in this paper, where $\sigma$ is substituted by $1/\Theta_s^2$ in the extreme ultrarelativistic limit, but the result of~\cite{Gruzinov:2014moa,Ciafaloni:2015xsr,Ciafaloni:2018uwe} has an extra logarithmic enhancement. This comes entirely from high-frequency gravitons, which seem to be irrelevant below the bound~\eqref{eq:KTb}. We plan to come back to this issue in a future work~\cite{tap1}.

Another interesting development is to investigate whether the eikonal operator has the coherent form of~\eqref{eq:Asr} at all frequencies, or non-linear corrections in the oscillators (in the exponent) are needed. The analysis of~\cite{Cristofoli:2021jas,Britto:2021pud} indicates that the first subleading correction to the waveform can be encoded in an eikonal operator that has the same functional form as~\eqref{eq:Asr}, at least below the threshold~\eqref{eq:KTb}. It will be important to clarify whether this is an all order property by studying both higher orders in the standard PM approach and the extreme ultrarelativistic limit of the first PM correction.

\subsection*{Acknowledgements} 

RR would like to thank IHES for hospitality during the final part of this work and Donato Bini and Thibault Damour for several enlightening discussions.
CH and RR acknowledge support by Institut Pascal at Université Paris-Saclay during the Paris-Saclay Astroparticle Symposium 2021, with the support of the P2IO Laboratory of Excellence (program “Investissements d’avenir” ANR-11-IDEX-0003-01 Paris-Saclay and ANR-10-LABX-0038), the P2I axis of the Graduate School Physics of Université Paris-Saclay, as well as IJCLab, CEA, IPhT, APPEC, the IN2P3 master projet UCMN and EuCAPT.
The research of RR is partially supported by the UK Science and Technology Facilities Council (STFC) Consolidated Grant ST/T000686/1. The research of CH (PDV) is fully (partially) supported by the Knut and Alice Wallenberg Foundation under grant KAW 2018.0116. Nordita is partially supported by Nordforsk.

\appendix

\section{Summary of the kinematics}
\label{app:5pkin}

Following closely Ref.~\cite{Parra-Martinez:2020dzs}, we write the $2\to 2$ kinematics in the Breit frame where the initial states have momenta
\begin{equation}
	\label{eq:inpie}
		p_1^\mu = -\overline{m}_1 u_1^\mu + \frac{q^\mu}{2} \;,\quad  p_2^\mu = -\overline{m}_2 u_2^\mu - \frac{q^\mu}{2} \;,
\end{equation}
where the quantities in bold are $(D-2)$ dimensional. The final states are
\begin{equation}
	\label{eq:inpfe}
	p_4^\mu = \overline{m}_1 u_1^\mu + \frac{q^\mu}{2} \;\quad  p_3^\mu = \overline{m}_2 u_2^\mu - \frac{q^\mu}{2} \;,
\end{equation}
so the momentum transferred $p_1+p_4=q$ is shared democratically between the in and out states and is orthogonal to the classical velocities $u_i$. We also introduce
\begin{equation}
	\label{eq:elconv}
	\overline{m}_i = \sqrt{m_i^2 + \frac{q^2}{4}} \;, \quad u_i^2=-1\;,\quad  y=- (u_1 u_2)= \frac{m_1 m_2 \sigma -\frac{q^2}{4}}{\overline{m}_1 \overline{m}_2}\;.
\end{equation}

Aligning the direction of classical motion along the $z$ axis, we can also introduce rapidity variables according to
\begin{equation}\label{}
		u_i^\mu = (\cosh y_i, \mathbf{0},\sinh y_i)\;, \quad q^\mu = (0,\mathbf{q},0)\;,
\end{equation}
and
\begin{equation}
	\label{eq:yicm}
	\begin{aligned}
		\cosh y_1 & = \frac{{\overline{m}}_1 +{\overline{m}}_2 y}{\sqrt{ {\overline{m}}_1^2+{\overline{m}}_2^2 +2 {\overline{m}}_1 {\overline{m}}_2 y}} =\frac{m_1}{\overline{m}_1}\frac{m_1 +m _2 \sigma}{\sqrt{ {m}_1^2+{m}_2^2 +2 {m}_1 {m}_2 \sigma}}  = \frac{E_1}{\overline{m}_1}\;, \\
		\cosh y_2 & =  \frac{{\overline{m}}_2 +{\overline{m}}_1 y}{\sqrt{ {\overline{m}}_1^2+{\overline{m}}_2^2 +2 {\overline{m}}_1 {\overline{m}}_2 y}}=\frac{m_2}{\overline{m}_2}\frac{m_2 +m _1 \sigma}{\sqrt{ {m}_1^2+{m}_2^2 +2 {m}_1 {m}_2 \sigma}} = \frac{E_2}{\overline{m}_2} \;, \\
		\sinh y_1 & = \frac{ {\overline{m}}_2 \sqrt{y^2 -1}}{\sqrt{ {\overline{m}}_1^2+{\overline{m}}_2^2 +2 {\overline{m}}_1 {\overline{m}}_2 y}} = \frac{p}{\overline{m}_1} \cos\frac{\Theta_s}{2}  \;,  \\
		\sinh y_2 & = -\frac{ {\overline{m}}_1 \sqrt{y^2 -1}}{\sqrt{ {\overline{m}}_1^2+{\overline{m}}_2^2 +2 {\overline{m}}_1 {\overline{m}}_2 y}} = - \frac{p}{\overline{m}_2} \cos\frac{\Theta_s}{2} \;,
	\end{aligned}
\end{equation}
where the spatial momentum $p$ is given by~\eqref{eq:ps} and the second formulation in each expression is derived by using~\eqref{eq:elconv}.

The direction of the radiation is of course defined by the momentum of the corresponding emitted soft particle. In $D=4$ we take
\begin{equation}
	k^\mu =(\omega, {\bf k}, k^L)= \omega n^\mu \;,\quad n^\mu =(1,\sin \theta \cos \phi, \sin \theta \sin \phi, \cos \theta)
	\label{GGG8}
\end{equation}
and introduce two orthogonal vectors $\epsilon_\phi k =\epsilon_\theta k=0$
\begin{equation}
	\label{eq:ephth}
	\epsilon_\phi^\mu= (0, -\sin \phi, \cos \phi, 0) \, \quad \epsilon_\theta^\mu = (0, \cos \theta \cos \phi, \cos \theta \sin \phi, - \sin \theta) \;.
\end{equation}
which can be used to define the physical polarisations of the graviton
\begin{equation}
	\label{eq:timesplus}
	\varepsilon^{\mu\nu}_\times = \frac{1}{2} (\epsilon^{\mu}_\phi \epsilon^{\nu}_\theta - \epsilon^{\nu}_\phi \epsilon^{\mu}_\theta)\;,\quad  \varepsilon^{\mu\nu}_+ = \frac{1}{2} (\epsilon^{\mu}_\theta \epsilon^{\nu}_\theta - \epsilon^{\mu}_\phi \epsilon^{\nu}_\phi)\;.
\end{equation}

\section{Spectrum of emitted momentum \texorpdfstring{$\frac{d\mathcal P^I_{\text{rad}}}{d\omega}$}{dPI/dw}} 
\label{app:dPI}

For completeness, let us briefly go back to the spatial components $\alpha= I$ of \eqref{Prad}, i.e.~the spectrum of emitted spatial momentum. They lead to the following angular integrals, which by rotational symmetry can be cast in the form
\begin{equation}\label{}
	I^I (p_n,p_m)
	= 
	\oint  \frac{\hat x^I}{(E_n-\vec k_n\cdot \hat x)(E_m-\vec k_m\cdot \hat x)}  
	\,\frac{d\Omega(\hat x)}{4\pi}
	=  
	a_{nm} \hat k_n^I + b_{nm} \hat k_m^I \,,
\end{equation}
so that
\begin{equation}\label{}
	a_{nm} = \frac{\hat k_n\cdot I - \cos\theta_{nm}\,\hat k_m\cdot I}{\sin^2\theta_{nm}}\,,\qquad
	b_{nm} = a_{mn}
	\,,
\end{equation}
where $\theta_{nm}$ is the angle between $\hat k_n$ and $\hat k_m$.
In terms of the rapidities
\begin{equation}\label{rapidities}
	E_{n,m} = m_{n,m} \cosh\psi_{n,m}\,,\qquad
	|\vec k_{n,m}| = m_{n,m}\sinh\psi_{n,m}\,,
\end{equation}
one finds
\begin{equation}\label{}
	\hat k_n\cdot I = \frac{1}{m_nm_m}\left[
	-\frac{\psi_m}{\sinh\psi_n\sinh\psi_m}
	-
	\frac{\operatorname{arctanh}f +\operatorname{arctanh}g}{2\sqrt{\sigma_{nm}^2-1}}
	\,\coth\psi_n
	\right],
\end{equation}
with
\begin{align}\label{}
	f(\psi_n,\psi_m,\sigma_{nm})
	&=
	\frac{2 e^{\psi_m}(e^{\psi_n}\sigma_{nm}^2-\sigma_{nm}\cosh\psi_m-\sinh\psi_n)}{(1+e^{2\psi_m}-2e^{\psi_n+\psi_m}\sigma_{nm})\sqrt{\sigma_{nm}^2-1}}\,,\\
	g(\psi_n,\psi_m,\sigma_{nm})
	&=
	\frac{2 e^{\psi_m}(\sigma_{nm}^2-\sigma_{nm} e^{\psi_n}\cosh\psi_m+e^{\psi_n}\sinh\psi_n)}{(e^{\psi_n}+e^{2\psi_m+\psi_n}-2e^{\psi_m}\sigma_{nm})\sqrt{\sigma_{nm}^2-1}}\,.
\end{align}
In the collinear limit $\theta_{nm}\to\pi^-$, this result simplifies to
\begin{equation}\label{simplermomentum}
	\hat k_n\cdot I  = \frac{\psi_n\coth\psi_n-\psi_m\coth\psi_m}{m_nm_m\sinh(\psi_n+\psi_m)}\,.
\end{equation}
The ZFL of the spectrum for the emission of spatial momentum is therefore given by
\begin{equation}\label{}
	\lim_{\omega\to0}\frac{d\mathcal P^I}{d\omega} 
	=
	\sum_{n,m} \frac{\kappa^2 \eta_n\eta_m}{(2\pi)^2}\,m_n^2m_m^2\left(\sigma_{nm}^2-\tfrac12\right)I^I(p_n,p_m)\,,
\end{equation}
for soft gravitons. Similar expressions can be obtained for vectors and scalar, following the same strategy adopted for $\operatorname{Im}2\delta$ and for $dE_{\text{rad}}/d\omega$ in the main body of the paper.

\providecommand{\href}[2]{#2}\begingroup\raggedright\endgroup

\end{document}